\pdfoutput=1
\documentclass[longauth]{aa}
\usepackage{natbib}
\usepackage{url,hyperref}
\usepackage{footmisc}
\usepackage{wrapfig,lipsum,booktabs}
\usepackage{txfonts}
\makeatletter
\renewcommand*{\@fnsymbol}[1]{\ifcase#1\or*\or$\dagger$\or$\ddagger$\or**\or$\dagger\dagger$\or$\ddagger\ddagger$\fi}
\makeatother
\usepackage{siunitx}
\usepackage{amssymb,amsmath}
\usepackage{makecell}
\usepackage[switch]{lineno}
\usepackage{amsmath}

\hypersetup{
        bookmarks=true,         % show bookmarks bar?
    unicode=false,          % non-Latin characters in AcrobatÕs bookmarks
    pdftoolbar=true,        % show AcrobatÕs toolbar?
    pdfmenubar=true,        % show AcrobatÕs menu?
    pdffitwindow=false,     % window fit to page when opened
    pdfstartview={FitH},    % fits the width of the page to the window
   pdfnewwindow=true,      % links in new window
    colorlinks=true,        % false: boxed links; true: colored links
   linkcolor=blue,          % color of internal links --> blue
      citecolor=blue,        % color of links to bibliography --> blue
      urlcolor=blue,           % color of external links --> blue      
}
\usepackage{ulem}

\begin{document}

\title{Constraining the cosmic-ray pressure in the inner Virgo Cluster using H.E.S.S. observations of M 87}
%\date{Received date / Accepted date }
\date{\today}

\author{H.E.S.S. Collaboration
\and F.~Aharonian \inst{\ref{DIAS},\ref{MPIK}}
\and F.~Ait~Benkhali \inst{\ref{LSW}}
\and C.~Arcaro\footnotemark[1] \inst{\ref{NWU}}
\and J.~Aschersleben \inst{\ref{Groningen}}
\and M.~Backes \inst{\ref{UNAM},\ref{NWU}}
\and V.~Barbosa~Martins\footnotemark[1] \inst{\ref{DESY}}
\and R.~Batzofin \inst{\ref{UP}}
\and Y.~Becherini \inst{\ref{APC},\ref{Linnaeus}}
\and D.~Berge \inst{\ref{DESY},\ref{HUB}}
\and K.~Bernl\"ohr \inst{\ref{MPIK}}
\and B.~Bi \inst{\ref{IAAT}}
\and M.~B\"ottcher \inst{\ref{NWU}}
\and C.~Boisson \inst{\ref{LUTH}}
\and J.~Bolmont \inst{\ref{LPNHE}}
\and J.~Borowska \inst{\ref{HUB}}
\and F.~Bradascio \inst{\ref{CEA}}
\and M.~Breuhaus \inst{\ref{MPIK}}
\and R.~Brose \inst{\ref{DIAS}}
\and F.~Brun \inst{\ref{CEA}}
\and B.~Bruno \inst{\ref{ECAP}}
\and T.~Bulik \inst{\ref{UWarsaw}}
\and C.~Burger-Scheidlin \inst{\ref{DIAS}}
\and T.~Bylund \inst{\ref{Linnaeus}}
\and S.~Caroff \inst{\ref{LAPP}}
\and S.~Casanova \inst{\ref{IFJPAN}}
\and R.~Cecil \inst{\ref{UHH}}
\and J.~Celic \inst{\ref{ECAP}}
\and M.~Cerruti \inst{\ref{APC}}
\and T.~Chand \inst{\ref{NWU}}
\and S.~Chandra \inst{\ref{NWU}}
\and A.~Chen \inst{\ref{Wits}}
\and J.~Chibueze \inst{\ref{NWU}}
\and O.~Chibueze \inst{\ref{NWU}}
\and G.~Cotter \inst{\ref{Oxford}}
\and J.~Damascene~Mbarubucyeye \inst{\ref{DESY}}
\and A.~Djannati-Ata\"i \inst{\ref{APC}}
\and K.~Egberts \inst{\ref{UP}}
\and J.-P.~Ernenwein \inst{\ref{CPPM}}
\and G.~Fichet~de~Clairfontaine \inst{\ref{LUTH}}
\and M.~Filipovic \inst{\ref{Sydney}}
\and G.~Fontaine \inst{\ref{LLR}}
\and M.~F\"u{\ss}ling \inst{\ref{DESY}}
\and S.~Funk \inst{\ref{ECAP}}
\and S.~Gabici \inst{\ref{APC}}
\and S.~Ghafourizadeh \inst{\ref{LSW}}
\and G.~Giavitto \inst{\ref{DESY}}
\and D.~Glawion \inst{\ref{ECAP}}
\and J.F.~Glicenstein \inst{\ref{CEA}}
\and P.~Goswami \inst{\ref{NWU}}
\and G.~Grolleron \inst{\ref{LPNHE}}
\and M.-H.~Grondin \inst{\ref{CENBG}}
\and L.~Haerer \inst{\ref{MPIK}}
\and M.~Haupt \inst{\ref{DESY}}
\and G.~Hermann \inst{\ref{MPIK}}
\and J.A.~Hinton \inst{\ref{MPIK}}
\and T.~L.~Holch \inst{\ref{DESY}}
\and D.~Horns \inst{\ref{UHH}}
\and M.~Jamrozy \inst{\ref{UJK}}
\and F.~Jankowsky \inst{\ref{LSW}}
\and V.~Joshi \inst{\ref{ECAP}}
\and I.~Jung-Richardt \inst{\ref{ECAP}}
\and E.~Kasai \inst{\ref{UNAM}}
\and K.~Katarzy{\'n}ski \inst{\ref{NCUT}}
\and R.~Khatoon \inst{\ref{NWU}}
\and B.~Kh\'elifi \inst{\ref{APC}}
\and W.~Klu\'{z}niak \inst{\ref{NCAC}}
\and Nu.~Komin \inst{\ref{Wits}}
\and K.~Kosack \inst{\ref{CEA}}
\and D.~Kostunin \inst{\ref{DESY}}
\and R.G.~Lang \inst{\ref{ECAP}}
\and S.~Le~Stum \inst{\ref{CPPM}}
\and F.~Leitl \inst{\ref{ECAP}}
\and A.~Lemi\`ere \inst{\ref{APC}}
\and M.~Lemoine-Goumard \inst{\ref{CENBG}}
\and J.-P.~Lenain \inst{\ref{LPNHE}}
\and F.~Leuschner \inst{\ref{IAAT}}
\and T.~Lohse \inst{\ref{HUB}}
\and A.~Luashvili \inst{\ref{LUTH}}
\and I.~Lypova \inst{\ref{LSW}}
\and J.~Mackey \inst{\ref{DIAS}}
\and D.~Malyshev \inst{\ref{IAAT}}
\and D.~Malyshev \inst{\ref{ECAP}}
\and V.~Marandon \inst{\ref{MPIK}}
\and P.~Marchegiani \inst{\ref{Wits}}
\and A.~Marcowith \inst{\ref{LUPM}}
\and P.~Marinos \inst{\ref{Adelaide}}
\and G.~Mart\'i-Devesa \inst{\ref{Innsbruck}}
\and R.~Marx \inst{\ref{LSW}}
\and M.~Meyer \inst{\ref{UHH}}
\and A.~Mitchell \inst{\ref{ECAP}}
\and R.~Moderski \inst{\ref{NCAC}}
\and L.~Mohrmann \inst{\ref{MPIK}}
\and A.~Montanari \inst{\ref{CEA}}
\and E.~Moulin \inst{\ref{CEA}}
\and J.~Muller \inst{\ref{LLR}}
\and K.~Nakashima \inst{\ref{ECAP}}
\and M.~de~Naurois\footnotemark[1] \inst{\ref{LLR}}
\and J.~Niemiec \inst{\ref{IFJPAN}}
\and A.~Priyana~Noel \inst{\ref{UJK}}
\and P.~O'Brien \inst{\ref{Leicester}}
\and S.~Ohm\footnotemark[1] \inst{\ref{DESY}}
\and L.~Olivera-Nieto \inst{\ref{MPIK}}
\and E.~de~Ona~Wilhelmi \inst{\ref{DESY}}
\and S.~Panny \inst{\ref{Innsbruck}}
\and M.~Panter \inst{\ref{MPIK}}
\and R.D.~Parsons \inst{\ref{HUB}}
\and G.~Peron \inst{\ref{APC}}
\and S.~Pita \inst{\ref{APC}}
\and D.A.~Prokhorov \inst{\ref{Amsterdam}}
\and H.~Prokoph \inst{\ref{DESY}}
\and G.~P\"uhlhofer \inst{\ref{IAAT}}
\and A.~Quirrenbach \inst{\ref{LSW}}
\and P.~Reichherzer \inst{\ref{CEA}}
\and A.~Reimer \inst{\ref{Innsbruck}}
\and O.~Reimer \inst{\ref{Innsbruck}}
\and M.~Renaud \inst{\ref{LUPM}}
\and F.~Rieger \inst{\ref{MPIK}}
\and G.~Rowell \inst{\ref{Adelaide}}
\and B.~Rudak \inst{\ref{NCAC}}
\and E.~Ruiz-Velasco \inst{\ref{MPIK}}
\and V.~Sahakian \inst{\ref{Yerevan}}
\and H.~Salzmann \inst{\ref{IAAT}}
\and D.A.~Sanchez \inst{\ref{LAPP}}
\and A.~Santangelo \inst{\ref{IAAT}}
\and M.~Sasaki \inst{\ref{ECAP}}
\and J.~Sch\"afer \inst{\ref{ECAP}}
\and F.~Sch\"ussler \inst{\ref{CEA}}
\and U.~Schwanke \inst{\ref{HUB}}
\and J.N.S.~Shapopi \inst{\ref{UNAM}}
\and H.~Sol \inst{\ref{LUTH}}
\and A.~Specovius \inst{\ref{ECAP}}
\and S.~Spencer \inst{\ref{ECAP}}
\and {\L.}~Stawarz \inst{\ref{UJK}}
\and R.~Steenkamp \inst{\ref{UNAM}}
\and S.~Steinmassl \inst{\ref{MPIK}}
\and C.~Steppa \inst{\ref{UP}}
\and I.~Sushch \inst{\ref{NWU}}
\and H.~Suzuki \inst{\ref{Konan}}
\and T.~Takahashi \inst{\ref{KAVLI}}
\and T.~Tanaka \inst{\ref{Konan}}
\and A.M.~Taylor \inst{\ref{DESY}}
\and R.~Terrier \inst{\ref{APC}}
\and M.~Tsirou \inst{\ref{DESY}}
\and N.~Tsuji \inst{\ref{RIKKEN}}
\and Y.~Uchiyama \inst{\ref{Rikkyo}}
\and C.~van~Eldik \inst{\ref{ECAP}}
\and B.~van~Soelen \inst{\ref{UFS}}
\and M.~Vecchi \inst{\ref{Groningen}}
\and J.~Veh \inst{\ref{ECAP}}
\and C.~Venter \inst{\ref{NWU}}
\and J.~Vink \inst{\ref{Amsterdam}}
\and T.~Wach \inst{\ref{ECAP}}
\and S.J.~Wagner \inst{\ref{LSW}}
\and R.~White \inst{\ref{MPIK}}
\and A.~Wierzcholska \inst{\ref{IFJPAN}}
\and Yu~Wun~Wong \inst{\ref{ECAP}}
\and M.~Zacharias \inst{\ref{LSW},\ref{NWU}}
\and D.~Zargaryan \inst{\ref{DIAS}}
\and A.A.~Zdziarski \inst{\ref{NCAC}}
\and A.~Zech \inst{\ref{LUTH}}
\and S.~Zouari \inst{\ref{APC}}
\and N.~\.Zywucka\footnotemark[1] \inst{\ref{NWU}}
}

\institute{
Dublin Institute for Advanced Studies, 31 Fitzwilliam Place, Dublin 2, Ireland \label{DIAS} \and
Max-Planck-Institut f\"ur Kernphysik, P.O. Box 103980, D 69029 Heidelberg, Germany \label{MPIK} \and
Landessternwarte, Universit\"at Heidelberg, K\"onigstuhl, D 69117 Heidelberg, Germany \label{LSW} \and
Kapteyn Astronomical Institute, University of Groningen, Landleven 12, 9747 AD Groningen, The Netherlands \label{Groningen} \and
University of Namibia, Department of Physics, Private Bag 13301, Windhoek 10005, Namibia \label{UNAM} \and
Centre for Space Research, North-West University, Potchefstroom 2520, South Africa \label{NWU} \and
DESY, D-15738 Zeuthen, Germany \label{DESY} \and
Institut f\"ur Physik und Astronomie, Universit\"at Potsdam,  Karl-Liebknecht-Strasse 24/25, D 14476 Potsdam, Germany \label{UP} \and
Université de Paris, CNRS, Astroparticule et Cosmologie, F-75013 Paris, France \label{APC} \and
Department of Physics and Electrical Engineering, Linnaeus University,  351 95 V\"axj\"o, Sweden \label{Linnaeus} \and
Institut f\"ur Physik, Humboldt-Universit\"at zu Berlin, Newtonstr. 15, D 12489 Berlin, Germany \label{HUB} \and
Institut f\"ur Astronomie und Astrophysik, Universit\"at T\"ubingen, Sand 1, D 72076 T\"ubingen, Germany \label{IAAT} \and
Laboratoire Univers et Théories, Observatoire de Paris, Université PSL, CNRS, Université de Paris, 92190 Meudon, France \label{LUTH} \and
Sorbonne Universit\'e, Universit\'e Paris Diderot, Sorbonne Paris Cit\'e, CNRS/IN2P3, Laboratoire de Physique Nucl\'eaire et de Hautes Energies, LPNHE, 4 Place Jussieu, F-75252 Paris, France \label{LPNHE} \and
IRFU, CEA, Universit\'e Paris-Saclay, F-91191 Gif-sur-Yvette, France \label{CEA} \and
Friedrich-Alexander-Universit\"at Erlangen-N\"urnberg, Erlangen Centre for Astroparticle Physics, Erwin-Rommel-Str. 1, D 91058 Erlangen, Germany \label{ECAP} \and
Astronomical Observatory, The University of Warsaw, Al. Ujazdowskie 4, 00-478 Warsaw, Poland \label{UWarsaw} \and
Université Savoie Mont Blanc, CNRS, Laboratoire d'Annecy de Physique des Particules - IN2P3, 74000 Annecy, France \label{LAPP} \and
Instytut Fizyki J\c{a}drowej PAN, ul. Radzikowskiego 152, 31-342 Krak{\'o}w, Poland \label{IFJPAN} \and
Universit\"at Hamburg, Institut f\"ur Experimentalphysik, Luruper Chaussee 149, D 22761 Hamburg, Germany \label{UHH} \and
School of Physics, University of the Witwatersrand, 1 Jan Smuts Avenue, Braamfontein, Johannesburg, 2050 South Africa \label{Wits} \and
University of Oxford, Department of Physics, Denys Wilkinson Building, Keble Road, Oxford OX1 3RH, UK \label{Oxford} \and
Aix Marseille Universit\'e, CNRS/IN2P3, CPPM, Marseille, France \label{CPPM} \and
School of Science, Western Sydney University, Locked Bag 1797, Penrith South DC, NSW 2751, Australia \label{Sydney} \and
Laboratoire Leprince-Ringuet, École Polytechnique, CNRS, Institut Polytechnique de Paris, F-91128 Palaiseau, France \label{LLR} \and
Universit\'e Bordeaux, CNRS, LP2I Bordeaux, UMR 5797, F-33170 Gradignan, France \label{CENBG} \and
Obserwatorium Astronomiczne, Uniwersytet Jagiello{\'n}ski, ul. Orla 171, 30-244 Krak{\'o}w, Poland \label{UJK} \and
Institute of Astronomy, Faculty of Physics, Astronomy and Informatics, Nicolaus Copernicus University,  Grudziadzka 5, 87-100 Torun, Poland \label{NCUT} \and
Nicolaus Copernicus Astronomical Center, Polish Academy of Sciences, ul. Bartycka 18, 00-716 Warsaw, Poland \label{NCAC} \and
Laboratoire Univers et Particules de Montpellier, Universit\'e Montpellier, CNRS/IN2P3,  CC 72, Place Eug\`ene Bataillon, F-34095 Montpellier Cedex 5, France \label{LUPM} \and
School of Physical Sciences, University of Adelaide, Adelaide 5005, Australia \label{Adelaide} \and
Leopold-Franzens-Universit\"at Innsbruck, Institut f\"ur Astro- und Teilchenphysik, A-6020 Innsbruck, Austria \label{Innsbruck} \and
Department of Physics and Astronomy, The University of Leicester, University Road, Leicester, LE1 7RH, United Kingdom \label{Leicester} \and
GRAPPA, Anton Pannekoek Institute for Astronomy, University of Amsterdam,  Science Park 904, 1098 XH Amsterdam, The Netherlands \label{Amsterdam} \and
Yerevan Physics Institute, 2 Alikhanian Brothers St., 375036 Yerevan, Armenia \label{Yerevan} \and
Department of Physics, Konan University, 8-9-1 Okamoto, Higashinada, Kobe, Hyogo 658-8501, Japan \label{Konan} \and
Kavli Institute for the Physics and Mathematics of the Universe (WPI), The University of Tokyo Institutes for Advanced Study (UTIAS), The University of Tokyo, 5-1-5 Kashiwa-no-Ha, Kashiwa, Chiba, 277-8583, Japan \label{KAVLI} \and
RIKEN, 2-1 Hirosawa, Wako, Saitama 351-0198, Japan \label{RIKKEN} \and
Department of Physics, Rikkyo University, 3-34-1 Nishi-Ikebukuro, Toshima-ku, Tokyo 171-8501, Japan \label{Rikkyo} \and
Department of Physics, University of the Free State,  PO Box 339, Bloemfontein 9300, South Africa \label{UFS}
}

\offprints{H.E.S.S.~collaboration,
\protect\\\email{\href{mailto:contact.hess@hess-experiment.eu}{contact.hess@hess-experiment.eu}};
\protect\\\protect\footnotemark[1] Corresponding authors
}

\abstract
{The origin of the gamma-ray emission from M\,87 is currently a matter of debate. This work aims to localize the very-high-energy (VHE; 100$\,$GeV - 100$\,$TeV) gamma-ray emission from M\,87 and probe a potential extended hadronic emission component in the inner Virgo Cluster. The search for a steady and extended gamma-ray signal around M\,87 can constrain the cosmic-ray energy density and the pressure exerted by the cosmic rays onto the intra-cluster medium (ICM), and allow us to investigate the role of the cosmic rays in the active galactic nucleus feedback as a heating mechanism in the Virgo Cluster. The High Energy Stereoscopic System (H.E.S.S.) telescopes are sensitive to VHE gamma rays and have been utilized to observe M\,87 since 2004. We utilized a Bayesian block analysis to identify M\,87 emission states with H.E.S.S. observations from 2004 until 2021, dividing them into low, intermediate, and high states. Because of the causality argument, an extended ($\gtrsim$kpc) signal is allowed only in steady emission states. Hence, we fitted the morphology of the 120$\,$h low state data and found no significant gamma-ray extension. Therefore, we derived for the low state an upper limit of $\ang{;;58}$(corresponding to $\approx$$4.6\,$kpc) in the extension of a single-component morphological model described by a rotationally symmetric 2D Gaussian model at 99.7\% confidence level. Our results exclude the radio lobes ($\approx$30$\,$kpc) as the principal component of the VHE gamma-ray emission from the low state of M\,87. The gamma-ray emission is compatible with a single emission region at the radio core of M\,87. These results, with the help of two multiple-component models, constrain the maximum cosmic-ray to thermal pressure ratio $X_{\mathrm{CR,max.}}$$\lesssim$$0.32$ and the total energy in cosmic-ray protons (CRp) to $U_\mathrm{CR}$$\lesssim$5$\times10^{58}\,$erg in the inner 20$\,$kpc of the Virgo Cluster for an assumed CRp power-law distribution in momentum with spectral index $\alpha_\mathrm{p}$=2.1.}


\authorrunning{H.E.S.S. Collaboration et al.}

\titlerunning{The cosmic-ray pressure in the inner Virgo Cluster}

\maketitle

\makeatletter
\renewcommand*{\@fnsymbol}[1]{\ifcase#1\@arabic{#1}\fi}
\makeatother

\section{Introduction}
\label{sec:introduction}
Galaxy clusters are the largest gravitationally bound structures in the Universe. The Virgo Cluster is a massive cluster of galaxies centered around the radio galaxy M\,87 at 16.5$\,$Mpc \citep{M87dist,M87dist2} away from Earth and extends up to $\approx$$\ang{2.3;;}$ from its center \citep[$r_\mathrm{500}\approx$$662.6\,$kpc, ][]{Simionescu2017}. It is known as a cool core (CC) cluster, where its central region
\citep[$\lesssim$0.01$r_\mathrm{180}$, ][]{coolingflow_review, Urban2011} is filled with a plasma that is colder and denser than the surrounding gas. The cooling flow model has been proposed to explain the formation of CC clusters based on the inward flow of radiatively cooled material which, in the absence of heating mechanisms, results in a mass deposition up to $\sim$1000$\,\mathrm{M_{\odot}}\,\mathrm{yr^{-1}}$. In contrast to this prediction, one to two orders of magnitude lower mass deposition rates are observed in CC clusters, which results in the rise of the cooling flow problem. The discrepancy points to the need for an additional heating mechanism to counterbalance the radiative cooling of the intra-cluster material (ICM) \citep{cooling_flow_theory,corecluster1,corecluster2}.

Thermal conduction is not capable of solely heating the ICM, since the necessary ICM conductivity would exceed the theoretical maximum, the Spitzer conductivity \citep{spitzer}. The dissipation of sound waves and turbulent motions have also been proposed as heating sources, though their contribution is likely not sufficient to account for the missing heating source \citep{Ruszkowski2004, Zhuravleva2014}. Furthermore, the mechanical heating by hot bubbles \citep{Bruggen2002, Mathews2006} can also contribute to the ICM heating, though the energy available is limited due to the disruption of the bubbles \citep{Pfrommer2013}.

In addition to the aforementioned feedback channels, the active galactic nucleus (AGN) feedback via cosmic rays \citep[CR,][]{cooling_flow_theory,coolingflow_review,Guo2008} accelerated in the central AGN region could provide the necessary heat to prevent the ICM from further cooling down. \citet{corecluster1,corecluster2} propose that the heating by the AGN feedback and the thermal conduction counterbalances the radiative cooling at any distance from the cluster center (steady-state system). The AGN accretes cooled gas and launches relativistic jets of particles. This process transfers energy to the surrounding gas and delays its radiative cooling. The CR electrons (CRe) injected by the AGN suffer severe energy losses via synchrotron and inverse Compton emission at GeV-TeV energies. In fact, 10$\;$GeV electrons in $\sim$10$\mu G$ magnetic fields, as in the M\,87 radio lobes \citep{LOFAR}, have a radiative lifetime of $\sim$10$\,$Myr \citep{CR_transport}. CR protons and nuclei (CRp), also accelerated in the central AGN, accumulate and fill the cluster over cosmic timescales due to the persistent infall of cooled gas and to the long radiative lifetimes of CRp. In reality, the radiative lifetime of CRp above 10$\,$GeV in the ICM is at least 60 times longer than the lifetime of CRe at any energy in the same ICM \citep[Fig.~2 from][]{CR_transport}. Hadronic interactions of the CRp with the ICM lead to the production of charged and neutral pions. While charged pions decay into electrons and positrons (hereafter secondary electrons), neutral pions decay into gamma rays. Secondary electrons are responsible for at least part of the extended radio halo emission seen around several galaxy clusters \citep{corecluster2}. In fact, cluster radio halos are the primary evidence for the existence of CRs in galaxy clusters \citep{CR_transport}. In dense environments ($\sim$10$^{-3}\,$-$\,$10$^{-1}\,$cm$^{-3}$) filled with CRp such as the central region of galaxy clusters, neutral pion decay is expected to produce a spatially extended and steady gamma-ray signal.

There had been many attempts to predict and observe diffuse GeV gamma-ray emission from galaxy clusters, for instance, \citet{prospects_cluster_gamma,Fermi-LAT_stacked_cluster,Prokhorov2014,Fermi_cluster,Fermi_cluster_extended}. \citet{Xi2018} were the first to claim the detection of GeV gamma rays from the Coma Cluster and \citet{Baghmanyan2022} were the first to claim that the gamma-ray signal from the Coma Cluster is extended at GeV energies. The extended GeV emission from the Coma Cluster is better fitted in a hadronic ICM scenario, although other models are also able to explain the data \citep{Coma_detection}. In the TeV regime, searches have been conducted for GeV gamma-ray emission from the Coma, Abell 496, Abell 85, and the Perseus galaxy clusters \citep{Abell_HESS,Coma_HESS,clusters_HESS2009,cluster_VERITAS,galaxyclusters,MAGIC_cluster} without success.

Detection of extended very-high-energy (VHE; 100$\,$GeV-100$\,$TeV) gamma-ray emission from clusters would not only establish a new class of VHE gamma-ray emitter but also support the hypothesis of the AGN feedback by streaming CR in CC clusters by revealing the total energy accelerated in CRp. The presence of buoyant rising bubbles in M\,87 indicates that AGN feedback plays a significant role in the Virgo Cluster \citep{bubbles}. Due to its active nucleus and its proximity to Earth, M\,87 is the best candidate for searching for steady and extended VHE gamma-ray emission from a galaxy cluster. However, the highly variable gamma-ray emission from its AGN \citep{2005flare,2008flare,2010flare} dominates over a potential steady emission component. This poses a challenge when it comes to accessing an underlying steady component \citep{pair_halos}. Nevertheless, only the cluster diffuse emission, that is, the emission that is of hadronic origin, should extend beyond the jet and fill the inner region of the cluster \citep{galaxyclusters}. Therefore, we aim at reducing the contribution of the variable gamma-ray emission to probe the cluster diffuse emission through an extended gamma-ray signal in the low state of M\,87.

Imaging atmospheric Cherenkov telescopes (IACTs) are sensitive to VHE gamma rays from astrophysical sources \citep{Hinton2009}. IACTs have shown their capabilities to detect extragalactic extended gamma-ray emission \citep{cenA} with the High Energy Stereoscopic System (H.E.S.S.). This study uses observations of M\,87 with H.E.S.S. from 2004 until 2021 to probe the origin and size of the gamma-ray emission in the low state of M\,87. We first tested for deviations from a single point source by fitting a point-like model and a rotationally symmetric 2D Gaussian model to the emission. Afterwards, we compared the morphology results to known features from the radio and X-ray emissions. Finally, we considered two scenarios for a multiple-component emission to interpret the results in terms of the cosmic-ray pressure in the inner Virgo Cluster.

This paper is organized as follows. In Sect.~\ref{sec:methods} we introduce the H.E.S.S. observations of M\,87 and the analysis methods. In Sect.~\ref{sec:results} we present the results, regarding the origin and morphology of M\,87 gamma-ray emission. In Sect.~\ref{sec:discussion} we interpret the results in terms of the CRp pressure in the inner Virgo Cluster and the total energy in CRp. Finally, in Sect.~\ref{sec:summary} we summarize the results and consider the prospects.

\section{Methods}
\label{sec:methods}

\subsection{H.E.S.S. experiment and data analysis}
\label{sec:hess_m87}
H.E.S.S. \citep{crabhess2006} is an array of five IACTs located in the Khomas Highland in Namibia, 1800$\,$m above sea level. The experiment has operated since 2003 with four 12$\,$m telescopes (CT\,1-4) and since 2012 with an additional 28$\,$m telescope (CT\,5), not utilized in this analysis. In 2017 a major upgrade of the cameras of the first four telescopes (CT$\,$1-4) improved the read-out and the stability of the system \citep{upgrade_cameras}.

M\,87 has been observed with H.E.S.S. since 2004 during several monitoring campaigns and in reaction to alerts of flaring activities. The data utilized in this work span from 2004 up to 2021 and were selected considering the following selection cuts: at least three telescopes participating in the observation, a maximum zenith angle of observation of 50$^\circ$, and an energy threshold of 300$\,$GeV for all the reconstructed events. These selection cuts ensure that the later reconstructed gamma-ray events have improved angular resolution. Since the majority of the observations are from the first phase of the experiment (CT$\,$1-4), CT$\,$5 data were not utilized in the analysis to avoid including different sources of systematic uncertainties. Finally, a total of 194$\,$h of observations was obtained after the application of the quality criteria suitable for spectral and morphological analyses \citep{crabhess2006}.

We analyzed the data with the H.E.S.S. analysis software applying the template-based reconstruction technique \citep[\textit{ImPACT};][]{ImPACT}. The gamma-hadron separation was performed through a multivariate analysis method \citep{boost_tree}. The \textit{Ring Background} and the \textit{Reflected-region Background} \citep{background_techniques} techniques provided an estimation of the remaining background. While the first method is ideal for studying the morphology of the emission, the second method is optimized for spectral studies. The point spread function (PSF), that is, the response of the instrument to a point-like source, was estimated based on Monte Carlo simulations that accounted for the observation conditions. With the template reconstruction, the PSF reaches a $68$\% containment radius of approximately $\ang{0.05;;}$ for energies above $1\,$TeV. We estimated the flux and spectrum of M\,87 from a region within $\theta$$\leq$$\ang{0.071;;}$ around the radio core of M\,87, that is, where the supermassive black hole (SMBH) is located. The exact position of the M\,87 radio core is at right ascension (RA) 12h30m49.423s and declination (DEC) \ang{12;23;28.04} \citep{M87position_2}. The region is defined to optimize the telescope's sensitivity toward a point-like source \citep{ImPACT}. A forward-folding method \citep{Piron2001} yielded the best power-law (PL) spectrum that fits the data above a safe energy threshold \citep{crabhess2006}.

\subsection{The light curve and Bayesian blocks}
\label{sec:bayesian_methods}
Based on the gamma-ray flux derived for the individual observations, we binned the flux data points in time, weighting them according to their statistical uncertainties. The result is the long-term light curve of the source.

To identify the low flux state, we utilized a Bayesian blocks algorithm \citep{bayesian0,bayesian} on the 30-day-binned light curve. We also investigated other bin sizes for the light curve; these are shown in Appendix~\ref{app:lightcurve}. The Bayesian block analysis is a method for detecting statistically significant changes in data from counting detectors and can be utilized to estimate flux levels assuming a piecewise-constant representation of time series data. It is effective also for non-constant sampling rates and hence is a very useful method for VHE gamma-ray astronomy \citep{Bayesian_VHE}. The method requires setting a prior function that directly influences the expected total number of blocks. We chose the prior function such that the false-positive rate, that is the chance of wrongly detecting a flux change, is at the level of 5\%
\citep{bayesian}. Studies that directly rely on the data points to define the low state, that is, without the use of the Bayesian blocks, cannot assure a low false-positive rate, leading to an unreliable low-state data set.

\begin{figure}[t!]
\centering
\includegraphics[width=\textwidth/2]{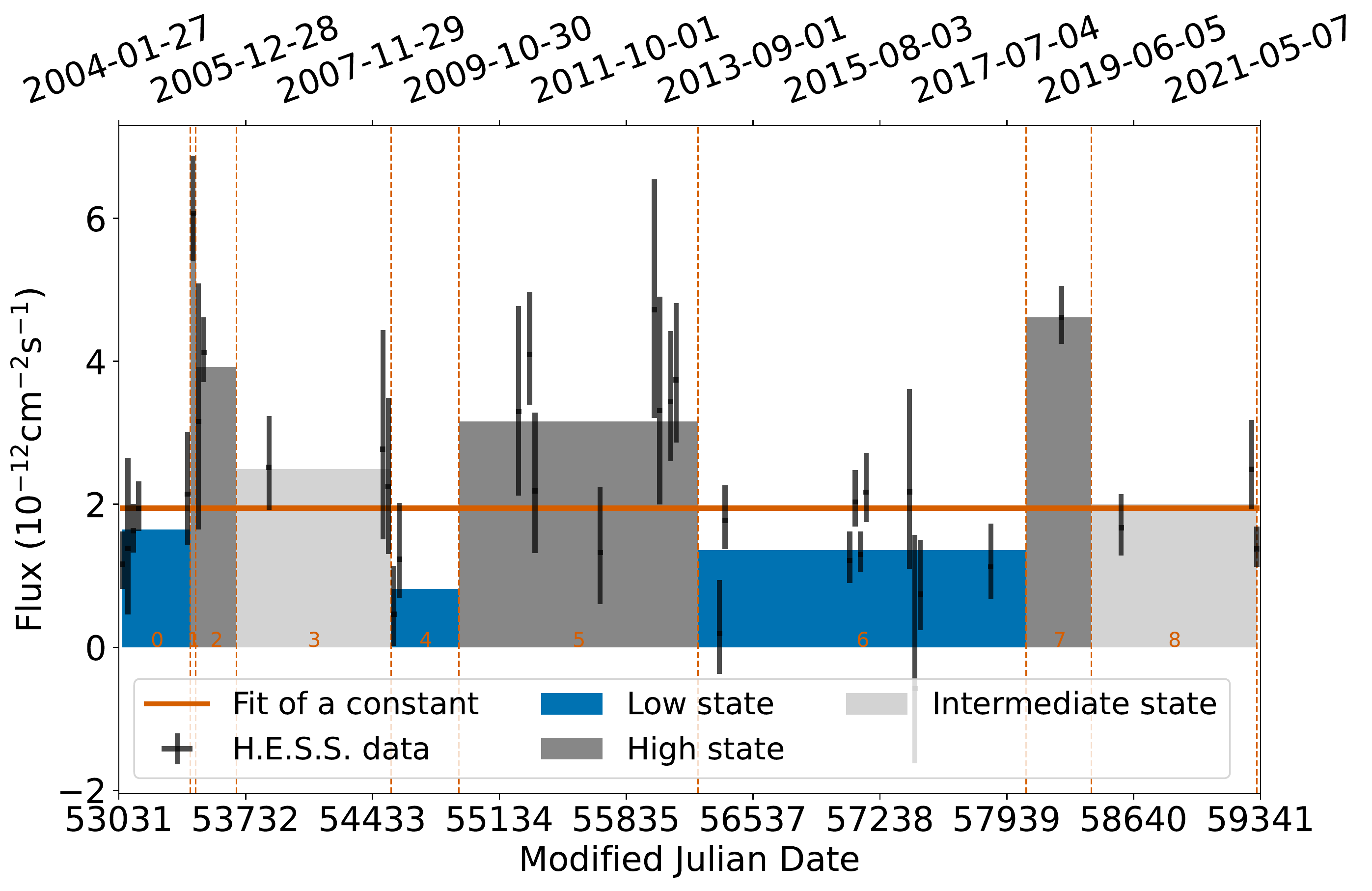}
\caption[H.E.S.S. light curve with its Bayesian blocks.]{Thirty-day binned H.E.S.S. light curve above 300$\,$GeV with its Bayesian blocks and source states. The H.E.S.S. data points are displayed by the black data points and a constant fit by the orange solid line. Dashed orange lines indicate a change in the flux level. The blue blocks represent the low state, the light gray the intermediate and the dark gray blocks represent the high state blocks. The orange numbers at the bottom indicate the labels of the blocks.}
\label{fig:bayesian}
\end{figure}

To assign the Bayesian blocks to distinct source states, we inspected Fig. \ref{fig:bayesian} and identified that Block 3 contains the 2008 VHE gamma-ray flare \citep{2008flare}. Furthermore, block 8 contains one data point more than 1$\,\sigma$ above the average flux. Therefore, blocks 3 and 8 are not allowed in the low state, which was then defined as the blocks below the average flux. Blocks 3 and 8 were assigned to the intermediate state since they are slightly (<30\%) higher than the average flux. The remaining blocks were assigned to the high state. Dividing the Bayesian blocks into three states helps us achieve a clear distinction between the low and high-state data sets, which minimizes the variable emission present in the low state. The definition of source states utilized in this study is focused on retrieving a low-state data set from the 30-day binned light curve with a minimum contribution from a variable component and might not be representative for flaring studies. 

The low state is our prime target for probing extended gamma-ray emission in M\,87 data. In the low state, the contribution from the tails of the PSF distribution from a point-like component at the (variable) core of M\,87 is minimized. Therefore, H.E.S.S. is most sensitive to detecting a diffuse emission from the inner Virgo Cluster in the low state. In addition to probing for extended emission in the low state of M\,87, we also tested if the source states have compatible emission regions through the morphology fit (Sect. \ref{sec:fit_algorithm}).

Since the choice of the light curve bin size also influences the defined source states, we also analyzed the data and fitted the morphology of a low state derived from a daily-binned light curve. The results were rather poor due to the lack of event statistics in the low state ($\approx$24$\,$h lifetime of observations). This test and the results shown in Appendix~\ref{app:lightcurve} favor the choice of a bin size larger than 15 days. We decided on a light curve with a bin size of 30 days since it provides good event statistics in the low state for a morphology fit, that is, the statistical uncertainties are at the same level as the systematic uncertainties in the position determination of $\ang{;;20}$ \citep{GC_localization}. Therefore, by binning the flux points in time we intentionally allowed a certain contribution from variable emission into the low state in favor of larger event statistics. We later accounted for and estimated the contribution of a varying source component to the low state through the introduction of hybrid models (Sect.~\ref{sec:discussion}).

\subsection{The morphology fit procedure}
\label{sec:fit_algorithm}

After dividing the data set into three different source states, we analyzed the individual states separately with the H.E.S.S. analysis software (Sect.~\ref{sec:hess_m87}). We obtained 2D distributions (sky maps) for the estimated background, the exposure time of the observations, the H.E.S.S. PSF, and the detected events. The PSF was convolved with the H.E.S.S. systematic uncertainty of $\ang{;;20}$ \citep{GC_localization}, although we tested the effect of the convolution on the final results and determined that it is minor. The sky maps, derived using a pixel size of $\ang{;;36}$ centered at the M\,87 radio core, were included in the spatial model of the emission.

We utilized predefined spatial models for the distribution of the intrinsic gamma-ray emission: a point-like model and a rotationally symmetric 2D Gaussian model given by: 

\begin{equation}
    f(x, y)=A\cdot exp \left(- \dfrac{(x-x_\mathrm{0})^2+(y-y_\mathrm{0})^2}{2\sigma_\mathrm{G}^2}\right ),
    \label{eq:Gaussian}
\end{equation}

\noindent where x and y are variables, x$_0$ and y$_0$ are the center of the function, A is the amplitude, and $\sigma_\mathrm{G}$ is the extension (width). While the point-like model represents the emission limited to the core region, the centrally-peaked Gaussian model allows us to probe the radiative cooling of the CRs during their transport from the AGN core toward the ICM. Alternatively, extended models such as a top-hat function could also be utilized, although they are less physically motivated than a Gaussian distribution for the case of CRp accelerated in the central AGN. 

The region inside a circle of 0.5$^\circ$ radius from the radio core defines the region of interest in the M\,87 low state data, that is, the region considered in the fit. Since counts are sampled from Poisson distributions, the fit algorithm is based on the Cash statistic \citep{Cash1979}. Afterwards, the model that results in the smallest test statistic (TS) is determined to be the one that best describes the data \footnote{We utilized the following Python packages throughout the analysis: NumPy 1.17.2 \citep{numpy}, SciPy 1.3.1 \citep{scipy}, Matplotlib 3.5.2 \citep{matplotlib}, Astropy 3.2.2 \citep{astropy:2013,astropy:2018}), Gammapy 0.17 \citep{gammapy:2017,cosimo}, and Sherpa 4.12.0 \citep{sherpa0,sherpa}.}.

A set of systematic checks confirms the stability of the results toward different configurations\footnote{More details in the PhD Thesis \citet{victor_phd}.}. We analyzed the M\,87 low state data several times under different circumstances: 1) with the maximum zenith angle of observation set at 45$^\circ$; 2) with the energy threshold set at 0.7$\,$TeV; 3) with a sky map bin size of $\ang{;;18}$ and 4) with a shift of $\ang{;;18}$ in the center of the sky map. The checks show that the results are stable against these changes. Furthermore, we reanalyzed the gamma-ray extension of the Crab Nebula \citep{crab} with the same procedure described here and obtained compatible (within 1 $\sigma$) results. 

A cross-check analysis with an independent analysis chain based on a semi-analytical shower model \citep{M++} also confirmed the robustness of the results of this work from the H.E.S.S. data analysis (Sect.~\ref{sec:hess_m87}) up to the morphology fit (Sect.~\ref{sec:fit_algorithm}).

\section{Results}
\label{sec:results}

In this section, we present the results of the analysis of the H.E.S.S. observations of M\,87 following the procedure described in Sect.~\ref{sec:methods}. First, we derive the light curve and source states in Sect.~\ref{sec:BB}. Afterwards, we present the results of the morphology fit of the low state in Sect.~\ref{sec:morphology_results} and derive conclusions about its origin in Sect.~\ref{sec:localization}.

\subsection{Bayesian blocks and source states}
\label{sec:BB}

\begin{table}[t!]
\caption[Bayesian blocks of the 30-day-binned H.E.S.S. light curve.]{The Bayesian blocks of the 30-day-binned H.E.S.S. light curve with its classification into low, intermediate, and high flux states.}
\begin{center}
\begin{tabular}{c c c c }
\hline\hline
Block & State & Start date & End date\\
\hline
0 & Low & 2004-02-16 & 2005-02-25 \\
1 & High & 2005-02-25 & 2005-03-27 \\
2 & High & 2005-03-27 & 2005-11-07 \\
3 & Intermediate & 2005-11-07 & 2008-03-11 \\
4 & Low & 2008-03-11 & 2009-03-21 \\
5 & High & 2009-03-21 & 2012-10-31 \\
6 & Low & 2012-10-31 & 2017-10-20 \\
7 & High & 2017-10-20 & 2018-10-15 \\
8 & Intermediate & 2018-10-15 & 2021-04-17 \\
\hline
\end{tabular}
\end{center}
\label{table:states}
\end{table}

\begin{table*}
\caption[Statistical results of the H.E.S.S. analysis for the low, intermediate, and high flux states.]{Results of the H.E.S.S. analysis for the low, intermediate, and high flux states. The integrated flux is given above 300$\,$ GeV.}
\centering
\begin{tabular}{c c c c c c c c}
\hline\hline
State & \thead{Excess \\(counts)} & \thead{Excess-to-\\background ratio} & \thead{Significance\\($\sigma$)}
& \thead{Livetime \\(h)} & \thead{Flux at $1\,$TeV \\ ($\mathrm{10^{-13}\,TeV^{-1}\,cm^{-2}\,s^{-1}}$)} & \thead{Spectral \\ index} & \thead{Integrated flux \\ ($\mathrm{10^{-12}\,cm^{-2}\,s^{-1}}$)} \\
\hline
Low & 593.6 & 0.49 & 15.6 & 120.4 & 3.4$\pm$0.2 & 2.63$\pm$0.09 & 1.50$\pm$0.13 \\
Intermediate & 198.8 & 0.68 & 10.5 & 28.5 & 4.1$\pm$0.9 & 2.36$\pm$0.10 & 1.57$\pm$0.20 \\
High & 397.1 & 1.34 & 19.4 & 29.0 & 10.4$\pm$0.5 & 2.25$\pm$0.05 & 3.76$\pm$0.22 \\
\hline
\end{tabular}
\label{table:statesmaster}
\end{table*}

We applied the Bayesian blocks technique to the 30-day-binned light curve as discussed in Sect.~\ref{sec:bayesian_methods}. Figure~\ref{fig:bayesian} shows the M\,87 light curve with the derived Bayesian blocks and source states. The start and end dates of the blocks are given in Table \ref{table:states} and the average gamma-ray flux above 300$\,$GeV throughout the entire period is $\approx$1.8$\times10^{-12}\,\mathrm{cm^{-2}}\,\mathrm{s^{-1}}$.

We stacked data from the individual blocks according to the defined source states and analyzed them separately. The results are summarized in Table \ref{table:statesmaster}. The low state has the largest data set ($\approx$120$\,$h), while the high state has the highest significance of detection ($\approx$19$\sigma$). The spectral index of a PL energy distribution \footnote{$d\phi/ dE$=$\phi_0 (E/$TeV$)^{-\Gamma}$, where $\Gamma$ is the spectral index, $E$ the gamma-ray energy and $\phi_0$ the normalization at $1\,$TeV.} hardens from the low to the high state, consistent with previous results \citep{2005flare, MAGIC2020}.

\subsection{The morphology of the low state of M\,87}
\label{sec:morphology_results}

\begin{figure*}[ht!]
\centering
\includegraphics[width=\textwidth]{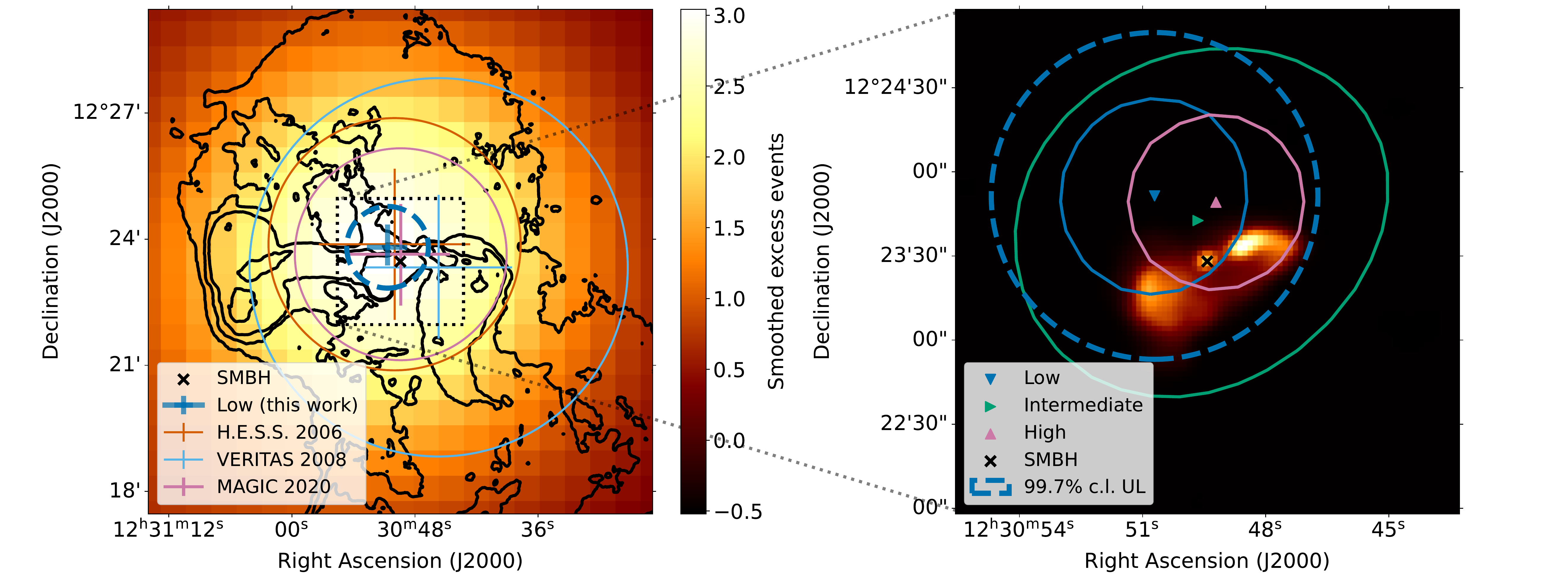}
\caption{(Left) H.E.S.S. low-state excess counts map with the derived extension UL plotted alongside previous ULs and the radio contour from VLA $\,90\,$cm map \citep{VLA, VLA90cm}. The H.E.S.S. map is smoothed with a rotationally symmetric 2D Gaussian function with $\sigma_\mathrm{G}=\ang{;;18}$ for better visualization. The best-fit position is shown by the blue marker for the Gaussian model with its $\sigma_\mathrm{G}$ extension UL at 99.7\% c.l given by the dashed dark blue circle. The blue error bars in the position include the 1$\,\sigma$ statistical uncertainty from the fit and $\ang{;;20}$ systematic uncertainty \citep{GC_localization}. Previous results by H.E.S.S. \citep{2005flare}, VERITAS \citep{VERITAS2008}, and MAGIC \citep{MAGIC2020} are given by the orange, light blue, and purple circles, respectively. The radio core of M\,87 is marked by the black cross \citep{M87position_2}. The dotted black square indicates the region depicted in the right figure. (Right) The VLA $21\,$cm \citep{FIRST, VLA25} radio map of M\,87 with the best-fit position for the point-like model for low, intermediate, and high flux states given by the blue, green, and pink triangles, respectively. The solid line contours represent $3\,\sigma$ statistical uncertainties for the respective source states. The blue dashed circle shows the resulting 99.7\% c.l. of the Gaussian $\sigma_\mathrm{G}$ extension UL in the low state. The best-fit positions of the source states are consistent with each other and with the M\,87 radio core, and they agree with previous results.}
\label{fig:lowhighradio}
\end{figure*}

We performed a morphology fit to the M\,87 low-state gamma-ray emission to probe its extension. No indication of a spatially extended emission component is found when comparing a point-like and an extended model as described in Section \ref{sec:fit_algorithm}. An upper limit (UL) on the Gaussian $\sigma_\mathrm{G}$ extension of $\ang{;;58}$ was derived at 99.7\% confidence level (c.l.). This translates to a physical extension limit of $\approx$4.6$\,$kpc, assuming a distance to M\,87 of $\approx$16.5\,Mpc \citep{M87dist,M87dist2}. This result improves by a factor of approximately two over the latest results \citep{MAGIC2020}. Previous measurements for the best-fit position \citep{2005flare,VERITAS2008} are in agreement with our results as shown in Fig.~\ref{fig:lowhighradio} (left).

\subsection{The origin of the gamma-ray emission of M\,87}
\label{sec:localization}

\begin{table}[!ht]
\caption{Best-fit parameters of the morphology fit.}
\centering
\begin{tabular}{c c c c}
\hline\hline
Model & RA($\boldsymbol{^\circ}$) & DEC ($\boldsymbol{^\circ}$) & $\boldsymbol{\sigma_\mathrm{G}}$($\boldsymbol{^{\prime\prime}}$) \\
\hline
Point-like & 187.707$\pm$0.005 & \thead{12.398$\pm$0.003} & - \\
Gaussian & 187.711$\pm$0.002 & $12.397_{-0.002}^{+0.003}$ & $12_{-12}^{+15}$\\
\hline

\end{tabular}
\tablefoot{The best-fit parameters of the low state of M\,87 (see Table \ref{table:states}) considering the point-like and the rotationally symmetric 2D Gaussian models. 1$\,\sigma$ statistical errors are given. The lower limit in the $\sigma_\mathrm{G}$ does not reach the 1$\,\sigma$ level within the valid interval ($\sigma_\mathrm{G}$>0).}
\label{table:resultsextension}
\end{table}

The results of the best-fit position of a point-like and Gaussian model are given in Table \ref{table:resultsextension} for the low state. A slight shift of $\approx$$\ang{;;25}$ in the best-fit position of the point-like model from the radio core is present. To investigate this apparent shift, we derived the 3$\mathrm{\,\sigma}$ uncertainty contours. Figure~\ref{fig:lowhighradio} (right) shows the best-fit position of the point-like model with its 3$\mathrm{\,\sigma}$ statistical uncertainty contours and the 99.7\% c.l. extension UL of the low state. We also plot the Very Large Array (VLA) radio 21$\,$cm emission in color scale, since it traces the energetic electrons in the inner radio cocoon ($\lesssim$2$\,$kpc). The shift between the best-fit position of the low state and the radio core is less than 3$\,\sigma$ even without including the systematic uncertainty of $\ang{;;20}$. Therefore, the best-fit position of the point-like model is consistent with the radio core.

Our extension UL on the low state of M\,87 excludes the radio lobes ($\approx$30$\,$kpc; black contours in Fig.~\ref{fig:lowhighradio} left) as the principal component of the low state of M\,87 gamma-ray emission. Since the radio emission of the inner cocoon is still contained within the extension UL (Fig.~\ref{fig:lowhighradio} right), we conclude that the inner radio cocoon cannot be excluded as the principal component. Further observations of the low state of M\,87 will improve the extension UL and probe the region inside the inner radio cocoon.

\begin{figure*}[ht!]
\centering
\includegraphics[width=\textwidth]{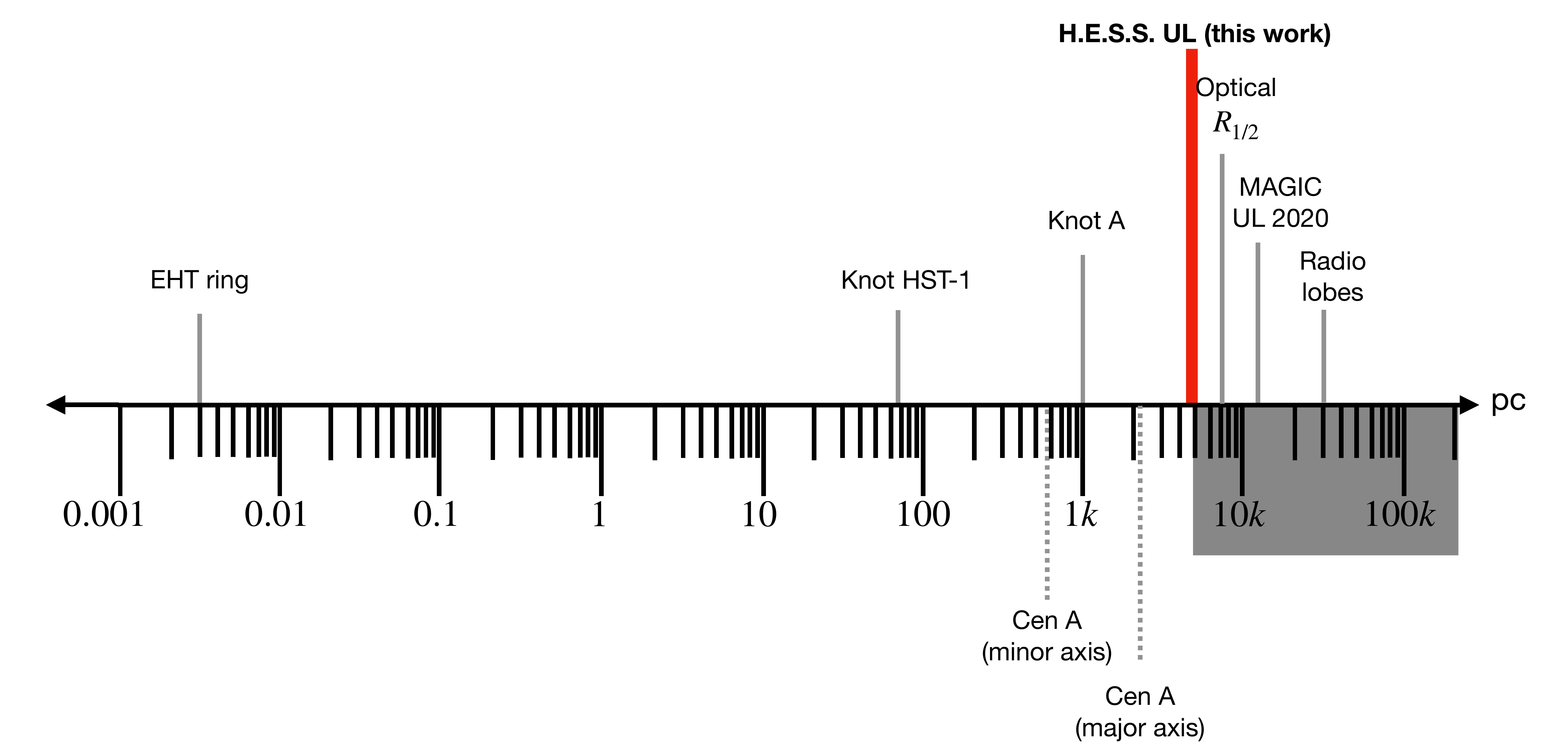}
\caption[Physical scales and prominent (jet) components in M\,87.]{Physical scales and prominent (jet) components in M\,87 in a logarithmic scale with the region excluded by this work shaded in gray. This work (Sect.~\ref{sec:morphology_results}) excludes the region $\gtrsim$4.6$\,$kpc as the origin of the VHE gamma-ray emission of the low state of M\,87. The half-light radius of the stars $R_{\mathrm{1/2}}$$\approx$7.2$\,$kpc \citep{M87_star_radius} and the dimensions of the radio lobes are also shown. The order of magnitude of the size of the EHT ring-like structure is shown \citep{EHTring, EHT2017}. The measured extension from Centaurus A is shown for comparison \citep{cenA}.}
\label{fig:UL_summary}
\end{figure*}

Since the sizes of the kpc-jet ($\lesssim$1$\,$kpc) and the X-ray knots are smaller than the H.E.S.S. extension UL, they could still contribute to (part of) M\,87 gamma-ray low-state emission. Furthermore, the results from \citet{EHT2017} show that the VHE emission during the low state of M\,87 cannot originate from a single zone leptonic scenario in the very close vicinity of the SMBH ($\lesssim$10$r_\mathrm{g}$$\approx$0.003$\,$pc, where $r_\mathrm{g}$ is the SMBH gravitational radius). On the other hand, the addition of a hadronic emission component in the close vicinity of the SMBH can indeed explain the broadband spectral emission \citep{Alfaro2022,Boughelilba2022, Xue2022}. Figure~\ref{fig:UL_summary} summarizes the scales of some known structures in M\,87, highlighting the region excluded by this work.

Despite the lack of an extension detected in the M\,87 low-state gamma-ray emission, we tested the hypothesis that the same emission region is responsible for the gamma rays in the different source states. Hence, we also fitted the morphology of the intermediate and high-flux states. None of the source states show significant extended emission and the best-fit positions of the point-like model of the M\,87 source states are consistent with each other.

\section{Discussion}

\label{sec:discussion}

Based on the extension UL derived in the previous section, we explore two plausible multiple-component scenarios that could lead to extended gamma-ray emission in the inner Virgo Cluster. Physical properties such as the energy content in CRp and the pressure they exert on the ICM are constrained and the role CRp play in feedback processes is discussed.

\subsection{Physical scenarios for an extended gamma-ray emission}
%pions
The M\,87 radio lobes extend up to $\approx$30$\,$kpc from the core and trace mildly relativistic electrons. The contribution from secondary electrons, that is resulting from the decay of charged pions, to M\,87 radio-lobe emission is subdominant \citep{Pfrommer2013}. Re-acceleration of electrons by plasma waves \citep{CR_transport} has been proposed to boost electrons with energies between 0.1 and 10$\,$GeV to tens of GeV, which would complement the radio-lobe emission. In fact, CRp would also be boosted by this re-acceleration mechanism. Given their longer radiative lifetime in the ICM, the CRp dominate the pressure in the inner Virgo Cluster over CRe. In general, reacceleration through plasma waves is a rather inefficient process and it is unlikely to boost electrons up to TeV energies beyond the kpc-jet region. Therefore, we can ignore the VHE gamma-ray contribution from reaccelerated mildly relativistic electrons in the cluster region. 

On the other hand, the streaming CRp might hadronically interact with the local target material and produce pions. The neutral pions could generate a detectable and extended VHE gamma-ray signal within the cluster region (Sect.~\ref{sec:introduction}). However, for this to occur, a dense target material and a strong CRp component must be present. The depletion of target material in the X-ray cavities \citep{Abdulla2019} could locally impede pion production, and subsequently reduce the total gamma-ray signal. While gamma-ray production through neutral pion decay is a promising scenario, it may face challenges due to the possible presence of material-depleted regions.

%leptonic emission regions
Particle acceleration could also take place in the lobes of radio galaxies as evidence from Fermi-LAT observations of Fornax~A indicates \citep{Ackermann2016}. The upscattering of the synchrotron and cosmic microwave background photons in the radio lobes by the locally accelerated CRe could also contribute to the VHE gamma-ray emission of M\,87 low state. As the H.E.S.S. extension UL derived here is significantly smaller than the size of the radio lobes, this process is very likely not the principal contributor to the low-state gamma-ray emission. On the other hand, electrons accelerated in the central AGN could significantly contribute to the gamma-ray signal through inverse Compton scattering in the photon fields of the core region. The cooling time of 10$\,$GeV CRe is typically $\sim$10$\,$Myr in a $\sim$10$\,\mu$G magnetic field and hence orders of magnitude shorter than the $p$-$p$ cooling time \citep[see Fig.~2 in][]{CR_transport}. Therefore, primary CRe are expected to mainly contribute to the emission in the close vicinity of the SMBH, which H.E.S.S. cannot resolve.
%kpc-jet emission
CRe accelerated in the central AGN lose most of their energy before reaching kpc distances, but could be reaccelerated via stochastic and/or shear particle acceleration to TeV energies \citep{Rieger2007}. These reaccelerated electrons would scatter via the inverse Compton process in the photon fields of the jet and produce an extended gamma-ray jet emission \citep{cenA}. The jet extension at radio to X-ray wavelengths is $\approx$1$\,$kpc ($\approx$$\ang{;;13}$), and also cannot be resolved by H.E.S.S. Therefore, a potential leptonic gamma-ray emission from the M\,87 kpc-jet will appear point-like and indistinguishable from the emission from the core in this work.

%pair halo and %p-gamma interaction
Gamma rays can also interact with extragalactic background light (EBL) photons on their way to Earth and produce electron-positrons pairs. These will initiate electromagnetic cascades, potentially producing an extended gamma-ray halo signal \citep{HESS_AGN_extension}. Given the close distance of M\,87 to Earth \citep[$\approx$16.5$\,$Mpc,][]{M87dist,M87dist2}, effects from the background light would become relevant to the gamma-ray spectrum at energies $\gtrsim$10$\,$TeV \citep{Franceschini2019}. In fact, only 9 events above the background level are reconstructed with energies above 10$\,$TeV in the low state, 4 in the intermediate state, and 13 in the high state. Therefore, the contribution of such an extended gamma-ray component in the low state of M\,87 can be safely neglected for morphological studies. 

The CRp from the jet could also interact with the local photon fields and produce neutral pions which could lead to an extended gamma-ray signal. However, \citet{Boughelilba2022} have shown that the accretion flow and the disk around the SMBH can be neglected as targets for particle-photon interactions in the jet.

%merger events
Another strong candidate for accelerating CRs and producing an extended gamma-ray signal is cluster mergers. Giant radio halos ($\gtrsim$200$\,$kpc) and relics \citep{CR_transport} are typically found in clusters with recent merging activities. Nevertheless, the Virgo Cluster does not have any of the aforementioned signals, although the cluster is also not yet completely dynamically relaxed 
\citep{Kashibadze2020}. During merger events, CRp are expected to be accelerated at shocks near the cluster's virial radius ($\approx$1.7$,$Mpc, that is, $\approx$6$^\circ$ for the Virgo Cluster)
\citep{Fermi_cluster_extended}, and they may be transported towards the cluster center, depending on the cluster's turbulent history. In the absence of turbulent advective transport that counterbalances diffusion, the CRp profile in galaxy clusters tends to flatten
\citep{CR_transport}. Hadronically interacting with the local ICM, these CRp can generate a diffuse gamma-ray signal up to the cluster's virial radius. However, H.E.S.S. would hardly detect such an extended gamma-ray signal, as its sensitivity degrades significantly to almost 80\% of the Crab flux for a source diameter of $\approx\ang{2;;}$ \citep{Casanova2008, Mitchell2022}.

In contrast to merger events, accretion events, that is, merging smaller virialized objects, might produce stronger shocks and accelerate CR more efficiently. \citet{Inoue_2005} have shown that for a $\sim\mu$G magnetic field in the shock region, CRp can be accelerated up to $\sim$10$^{18}\,$eV. These CRp would produce pairs of electron-positron, which would rapidly cool through inverse Compton scattering and synchrotron losses, ultimately generating a signal in the form of gamma rays with an energy flux of 10$^{-12}$-10$^{-11}\,$erg$\,$cm$^{-2}\,$s$^{-1}$, marginally detectable by H.E.S.S given the extension of the signal. For the Virgo Cluster the shock diameter can reach up to $\approx$$\ang{1.9;;}$ around the virial radius ($\approx$$\ang{6}$), that is, outside the region of interest of our analysis.

%DARK MATTER
Finally, the annihilation of hypothetical weakly interacting massive particles in the dark matter (DM) halo around M\,87 could also contribute to an extended gamma-ray signal, as investigated in \citet{Fermi_cluster_extended} at GeV energies. The size of the gamma-ray emission would strongly depend on the DM particle model. Therefore, for the sake of simplicity, the DM scenario is not covered in this study.

Among the discussed scenarios for an extended VHE gamma-ray signal in the low state of M\,87, we consider the neutral pion decay gamma-ray emission from $p$-$p$ interactions as the most likely one. To model the CR pressure and the gamma-ray emission from pion decay in the inner Virgo Cluster, the CRp energy and spatial distributions as well as the ICM density distribution have to be considered. Very little is known about CR in galaxy clusters and assumptions have to be made about their energy and spatial distribution. In contrast, the ICM density distribution can be estimated using X-ray measurements assuming an element composition of the plasma (Appendix~\ref{app:formalism}). However, X-ray measurements provide no information about the spatial distribution of the emission along the line of sight, which makes it challenging to construct a 3D model of the ICM without making further assumptions. Additionally, the X-ray surface brightness of the inner Virgo Cluster exhibits a complex morphology, featuring X-ray cavities that coincide with the inner radio lobes, as well as X-ray arcs with the matter at lower temperatures ($\approx$1$\,$keV) than the surrounding gas \citep{Chandra2002}. The composition of X-ray cavities and the nature of the sustaining pressure in radio galaxies are still not fully understood \citep{Abdulla2019}. Therefore, we adopt a simplified assumption of a radially symmetric ICM distribution, which does not account for the complex X-ray morphology of the cavities in the inner radio lobes. This approach may result in an overestimation of the gamma rays produced by neutral pion decay in the cavities if they are depleted of target material, which could ultimately affect the final upper limit on the CRp pressure. Furthermore, averaging the 2D X-ray brightness in azimuth leads to a 1$\,\sigma$ uncertainty on the radial X-ray brightness of approximately 85\% in the region close to the M\,87 core (within 4$\,$kpc) and approximately 30-40\% in the region up to $\ang{0.5;;}$ from the M\,87 core. Since these uncertainties propagate to the ICM density distribution, the limits derived in Sect.~\ref{sec:lofar} and Sect.~\ref{sec:steady} are approximations for the case of a radially symmetric ICM distribution.

We assume that CRp are distributed in momentum space according to a PL with spectral index $\alpha_\mathrm{p}$, which we vary from 2.1 to 2.6. PL distributions with spectral indices $\alpha_\mathrm{p}$$\gtrsim$2 are characteristic of first order Fermi acceleration \citep{Fermi1949, Rieger2007}. Due to the long cooling time of CRp in the ICM \citep{CR_transport}, no exponential cutoff is expected in the highest-energy part of the particle spectrum. The formalism utilized in this work to interpret the morphology results in terms of the CR to thermal pressure ratio is presented in Appendix~\ref{app:formalism} and is largely based on Appendix~B2 of \citet{corecluster2}. The parametrizations for the ICM density and temperature distributions in the Virgo Cluster shown in Appendix~\ref{app:formalism} are extracted from \citet{corecluster1,corecluster2}.

Finally, we considered two multiple-component models that lead to an extended gamma-ray component in the M\,87 low state: the magnetic confinement and the steady-state models. Both models are discussed and derived in Appendix~\ref{app:lofar} and Appendix~\ref{app:steady}, respectively. We defined hybrid models composed of one diffuse model and a point-like component to account for the leptonic emission from the AGN. The hybrid models have two free parameters: the amplitudes of the template and the point-like component. The center of the point-like component was fixed to the best-fit position of the pure point-like model, given in Table \ref{table:resultsextension}. In the following, we fitted the M\,87 low-state data to the hybrid models and discuss the implications of the results.

%Finally, we adopted two different approaches to estimate the CRp pressure in the inner Virgo Cluster: the LOFAR-based and the AGN feedback approaches.

\subsection{The magnetic confinement model}
\label{sec:lofar}

We fitted the hybrid model composed of the magnetic confinement model and a point-like source at the core of M\,87 to the H.E.S.S. data. The fit converged to a point-like model, showing that an additional diffuse component in the morphology model does not improve the fit. To place an UL on the gamma-ray flux from the diffuse component, we first defined a set of hybrid models with increasing fixed amplitude of the diffuse component. The new hybrid models have now only one free parameter accounting for the intensity of the point-like component. Finally, we fitted the H.E.S.S. data to this defined set of hybrid models and compared their TS with the TS of the fit of a pure point-like model. The maximum intensity of the diffuse component allowed by the morphology fit (at a 99.7\% c.l.) is found when the $\Delta \mathrm{TS}$ of the morphology fit reaches 3$\sigma$ preference for the point-like model in comparison to the hybrid model. The results of our analysis showed that the 3$\sigma$ preference is reached when the template component accounts for 45\% of the total low-state flux of M87. This translates into a maximum gamma-ray flux above 300$\,$GeV of $\lesssim$6.7$\times$10$^{-13}\,\mathrm{cm^{-2}s^{-1}}$ for an extended gamma-ray component in the magnetic confinement scenario. Likewise, the contribution of the point-like component is estimated to account for $\geq$55\% of the M\,87 low state emission.

To derive the maximum CR pressure allowed by the morphology fit of the hybrid model, the CRp distribution has to be taken into account. A centrally peaked CRp distribution is expected since we probe the source of CRp at the center of the cluster, that is, the AGN. Based on the equations from Appendix~\ref{app:formalism} \citep{corecluster1,corecluster2}, which describes the assumed ICM composition, density, and temperature distributions, we estimated the CRp distribution such that the CR to thermal pressure ratio ($X_\mathrm{CR}(r)$=$P_\mathrm{CR}(r)/P_\mathrm{th.}(r)$) is constant in the inner region of the cluster. A constant $X_\mathrm{CR}$ in the inner parts of the cluster is characteristic of an equilibrium state between the heating of the ICM by the streaming CRp and the ICM thermal cooling \citep{Pfrommer2013}. The results showed that the CR to thermal pressure ratio is $X_\mathrm{CR}$$\lesssim$$0.17$ for a proton distribution with $\alpha_\mathrm{p}$=$2.1$ at 99.7\% c.l. This estimate is directly influenced by the assumptions on the CRp and ICM spatial distributions. If large regions are material-depleted, as is possibly the case in the X-ray cavities \citep{Chandra2002}, the UL on the $X_\mathrm{CR}$ obtained in this study is underestimated, at least in the region of the cavities.

\begin{figure}
\centering
\includegraphics[width=\textwidth/2]{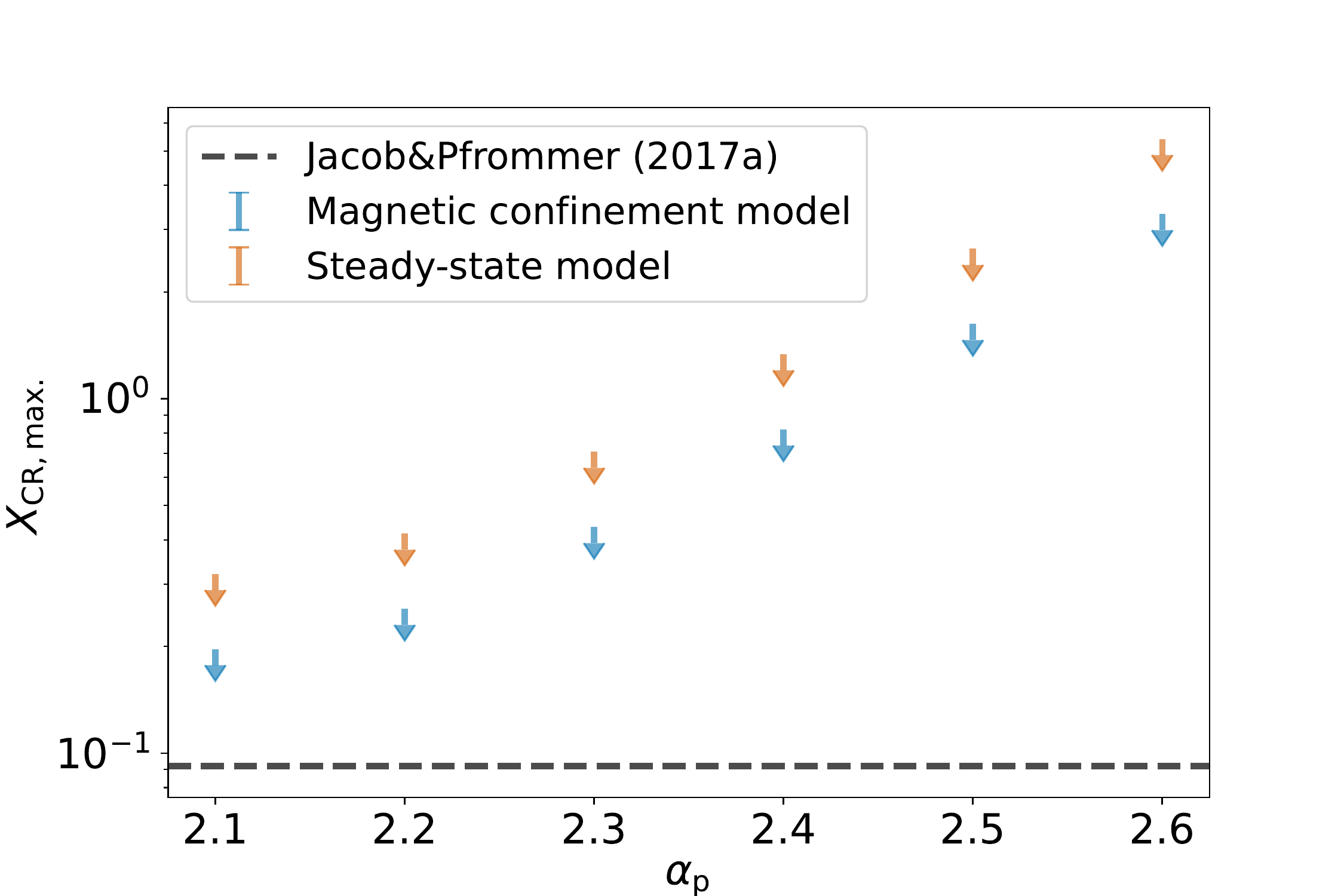}
\caption{The UL of the maximum CR to thermal pressure ratio $X_{\mathrm{CR,max.}}$=max($P_{\mathrm{CR}}(r)/P_{\mathrm{th.}}(r)$) within the inner 20$\,$kpc of the cluster for different proton spectral indices. Two different CRp spatial distributions were considered for a set of CRp spectral indices. The blue ULs are the result of the magnetic confinement approach (Sect.~\ref{sec:lofar}) and the orange ULs are the result of the steady-state approach (Sect.~\ref{sec:steady}). The prediction by \citet{corecluster1} is also shown in dashed black.}
\label{fig:XRC_complete}
\end{figure}

The CRp energy density e$_\mathrm{CR}$ was estimated and integrated within the volume to yield the total energy in CRp in the inner 20$\,$kpc of the Virgo Cluster of $U_\mathrm{CR}$$\lesssim$5$\times 10^{58}\,$erg. This is twice as much energy as the total energy estimated from a theoretical model consisting of shocks produced by outbursts that explains the M\,87 radio and X-ray emissions \citep{Forman2017}. \citet{Bruggen2002} has shown that the buoyant gas in a galaxy cluster can reach a distance of $\approx$20$\,$kpc after $\approx$15$\,$Myr. Therefore, for the streaming CRp to reach $U_\mathrm{CR}$$\lesssim$5$\times 10^{58}\,$erg in 15$\,$Myr, an average jet power of $P_\mathrm{j}$$\lesssim$6$\times 10^{43}\,$erg$\,\mathrm{s^{-1}}$ is necessary, considering the extreme case of 100\% efficiency in the CRp acceleration. Even though the estimates on $U_\mathrm{CR}$ and $P_\mathrm{j}$ are rather uncertain and model-dependent, the previous arguments show that these ULs are larger than (but of the order of magnitude of) the values expected from the literature \citep{Forman2017}. Furthermore, the UL on the $U_\mathrm{CR}$ derived in this work is twice as constraining as the result from \citet{Russia2022}, which placed an UL on $U_\mathrm{CR}$ up to 35$\,$kpc from the M\,87 core based on the \textit{Fermi} Large Area Telescope \citep[\textit{Fermi}-LAT; ][]{Abdollahi_2020} data. Our estimates above rely on the assumption that the spectral index of the CRp is $\alpha_\mathrm{p}$=2.1. Since the spectral index of the gamma rays from neutral pions is expected to follow the index of the CRp population, an additional brighter and steeper component in the M\,87 low-state emission would be necessary to account for the overall gamma-ray flux estimated with H.E.S.S. ($\alpha_\mathrm{\gamma}$=2.63$\pm$0.09; Table \ref{table:statesmaster}), possibly, the kpc-jet as in Centaurus A \citep{cenA}. For steeper CRp spectral indices our ULs on $X_\mathrm{CR}$ and $U_\mathrm{CR}$ become less constraining as seen in Fig.~\ref{fig:XRC_complete} by the blue ULs.

% Matthews2018: 8. 10^57 erg lobes M\,87
% Forman2017: 5. 10^57 erg (outburst)

\subsection{The steady-state model}
\label{sec:steady}

The fit of the M\,87 low-state data to the steady-state hybrid model converged into a pure point-like model. Nevertheless, by fitting a set of hybrid models with increasing fixed amplitude of the diffuse component, we derived a 99.7\% c.l. UL on the total flux of the diffuse component of $\lesssim$8.1$\times 10^{-13}\mathrm{cm^{-2}\,s^{-1}}$, that is $\lesssim$55\% of the low state of M\,87 VHE gamma-ray flux (Table \ref{table:statesmaster}). This UL is a factor of $\approx$4 larger than predicted by the model. Hence, the morphology fit was not able to probe the model proposed in \citet{corecluster1} for the Virgo Cluster. However, we could probe a scenario in which the gamma-ray distribution from neutral pion decay resembles the one from the steady-state model but is normalized to $\approx$55\% of the flux of M\,87 low state. Afterwards, we derived an UL on the maximum CR to thermal pressure ratio ($X_\mathrm{CR,max.}$) in the inner Virgo Cluster for the set of CRp spectral indices. The results are similar to the results of the first approach as shown by the orange ULs in Fig.~\ref{fig:XRC_complete}. The same approach as in Sect.~\ref{sec:lofar} is used here to account for the ICM and CRp spatial distributions. For a proton distribution with $\alpha_\mathrm{p}$=$2.1$, $X_\mathrm{CR,max.}$$\lesssim$$0.32$ and the total energy in CRp in the inner 20$\,$kpc of the Virgo Cluster is also constrained to $U_\mathrm{CR}$$\lesssim$5$\times10^{58}\,$erg. While the steady-state model predicts a level of $X_{\mathrm{CR,max.}}$$\approx$0.10 \citep[Fig.~A1 in ][]{corecluster1}, the H.E.S.S. UL is $X_{\mathrm{CR,max.}}$$\lesssim$$0.32$ for $\alpha_\mathrm{p}$=$2.1$ and becomes less constraining for steeper proton distributions, where the expected flux in VHE gamma rays is lower. Therefore, the H.E.S.S. $X_{\mathrm{CR,max.}}$ UL derived from the morphology fit of the M\,87 gamma-ray low state does not rule out the steady-state model regardless of the spectral index of the proton distribution.

\section{Summary and conclusions}
\label{sec:summary}
In this work, we aimed at localizing the VHE gamma-ray emission from M\,87, probing an extended emission in its low state, and testing its hadronic origin. Neutral pions are produced in $p$-$p$ interactions between the relativistic protons (CRp) from the jet and the ICM \citep{cooling_flow_theory,coolingflow_review}. The neutral pions decay almost immediately to gamma rays, which could be detected with H.E.S.S. as an extended and steady gamma-ray signal \citep{corecluster1,corecluster2}.

First, we investigated the VHE gamma-ray flux of M\,87 with H.E.S.S. between 2004 and 2021 and classified the source emission into low, intermediate, and high flux states based on a Bayesian block analysis \citep{bayesian0,bayesian, Bayesian_VHE}. We focused our studies on the low state since the detection of steady and extended emission could point to a hadronic origin and provide an estimate of the CRp pressure in the inner Virgo Cluster.

We did not detect extended emission via the morphology fit of the low state. Nevertheless, we derived an UL on the $\sigma_\mathrm{G}$ of a rotationally symmetric 2D Gaussian model of $\ang{;;58}$ ($\approx$4.6$\,$kpc) at 99.7\% c.l. The best-fit position of the point-like source model is compatible with the radio core (Fig.~\ref{fig:lowhighradio}) within 3$\sigma$ statistical uncertainty. Furthermore, our extension UL is twice as constraining as the latest result \citep{MAGIC2020} and, considering the uncertainties in the best-fit position, it excludes for the first time the M\,87 radio lobes \citep[$\approx$30$\,$kpc, ][]{VLA90cm} as the main contributor to 
the low state of M\,87 gamma-ray emission (Fig.~\ref{fig:lowhighradio} left). On the other hand, the inner radio cocoon (VLA 21$\,$cm), as shown in Fig.~\ref{fig:lowhighradio} right, cannot be ruled out as the principal component. Our UL on the extension lies already within the optical extent of M\,87 \citep[$R_{1/2}$$\approx$7.2$\,$kpc, ][]{M87_star_radius}. The origin of the VHE gamma-ray emission from the M\,87 source states are within the uncertainties consistent with a single origin at the M\,87 core.

We considered two plausible multiple-component scenarios to explain the gamma-ray emission in the low state of M\,87. We derived templates for the diffuse gamma-ray components of these two approaches and defined two hybrid models composed of a point-like component centered at M\,87 to account for the AGN emission and each of the respective diffuse templates. The morphology fit of the hybrid models allowed us to constrain the CRp pressure in the inner Virgo Cluster. The first approach (Appendix~\ref{app:lofar}) was based on the magnetic confinement model and the second approach (Appendix~\ref{app:steady}) was based on the steady-state model from \citet{corecluster1}. The results showed that the contribution of the diffuse component is constrained at 99.7\% c.l. to $\lesssim$45\% of the VHE gamma-ray flux detected from the low state of M\,87 for the magnetic confinement hybrid model and $\lesssim$55\% for the steady-state hybrid model. To interpret these limits in terms of CRp pressure in the inner Virgo Cluster, we first considered the CRp distributed as a PL in momentum with spectral index $\alpha_\mathrm{p}$ that we varied from 2.1 to 2.6. We utilized the formalism and the ICM parametrizations from \citet{corecluster1, corecluster2} for the ICM distribution to derive ULs on the maximum CR to thermal pressure ratio ($X_\mathrm{CR,max.}$) and the maximum energy in CRp ($U_\mathrm{CR}$) in the region up to 20$\,$kpc from the cluster center. For a CRp distribution with spectral index $\alpha_\mathrm{p}$=2.1, $X_\mathrm{CR,max.}$$\lesssim$$0.17$ for the hybrid models with the magnetic confinement template, while $X_\mathrm{CR,max.}$$\lesssim$$0.32$ for the hybrid model with the steady-state template. For steeper CRp distributions, the H.E.S.S. UL is less constraining. The energy in CRp, assuming $\alpha_\mathrm{p}$=2.1, is constrained at 99.7\% c.l. to $U_\mathrm{CR}$$\lesssim$5$\times10^{58}\,$erg up to 20$\,$kpc from M\,87 core in both approaches. This limit is larger than, but of the same order of magnitude of the total energy expected from a theoretical model consisting of shocks produced by outbursts from M\,87 \citep{Forman2017}. Our UL on the $U_\mathrm{CR}$ is also twice as constraining as the UL based on Fermi-LAT data \citep{Russia2022}.

\section*{Acknowledgments}
The support of the Namibian authorities and of the University of Namibia in facilitating the construction and operation of H.E.S.S. is gratefully acknowledged, as is the support by the German Ministry for Education and Research (BMBF), the Max Planck Society, the German Research Foundation (DFG), the Helmholtz Association, the Alexander von Humboldt Foundation, the French Ministry of Higher Education, Research and Innovation, the Centre National de la Recherche Scientifique (CNRS/IN2P3 and CNRS/INSU), the Commissariat à l’énergie atomique et aux énergies alternatives (CEA), the U.K. Science and Technology Facilities Council (STFC), the Irish Research Council (IRC) and the Science Foundation Ireland (SFI), the Knut and Alice Wallenberg Foundation, the Polish Ministry of Education and Science, agreement no. 2021/WK/06, the South African Department of Science and Technology and National Research Foundation, the University of Namibia, the National Commission on Research, Science \& Technology of Namibia (NCRST), the Austrian Federal Ministry of Education, Science and Research and the Austrian Science Fund (FWF), the Australian Research Council (ARC), the Japan Society for the Promotion of Science, the University of Amsterdam and the Science Committee of Armenia grant 21AG-1C085. We appreciate the excellent work of the technical support staff in Berlin, Zeuthen, Heidelberg, Palaiseau, Paris, Saclay, Tübingen and in Namibia in the construction and operation of the equipment. This work benefited from services provided by the H.E.S.S. Virtual Organisation, supported by the national resource providers of the EGI Federation. 

Finally, we thank De Gasperin who kindly provided the LOFAR radio emission of M\,87 at 140$\,$MHz \citep[Fig.~5 from ][]{LOFAR} and Pfrommer for the discussions on his model.

\bibliographystyle{aa}
\bibliography{references.bib}
\newpage

\begin{appendix}

\section{The long-term light curve of M\,87}
\label{app:lightcurve}

\begin{figure}
    \centering
    \includegraphics[width=\textwidth/2]{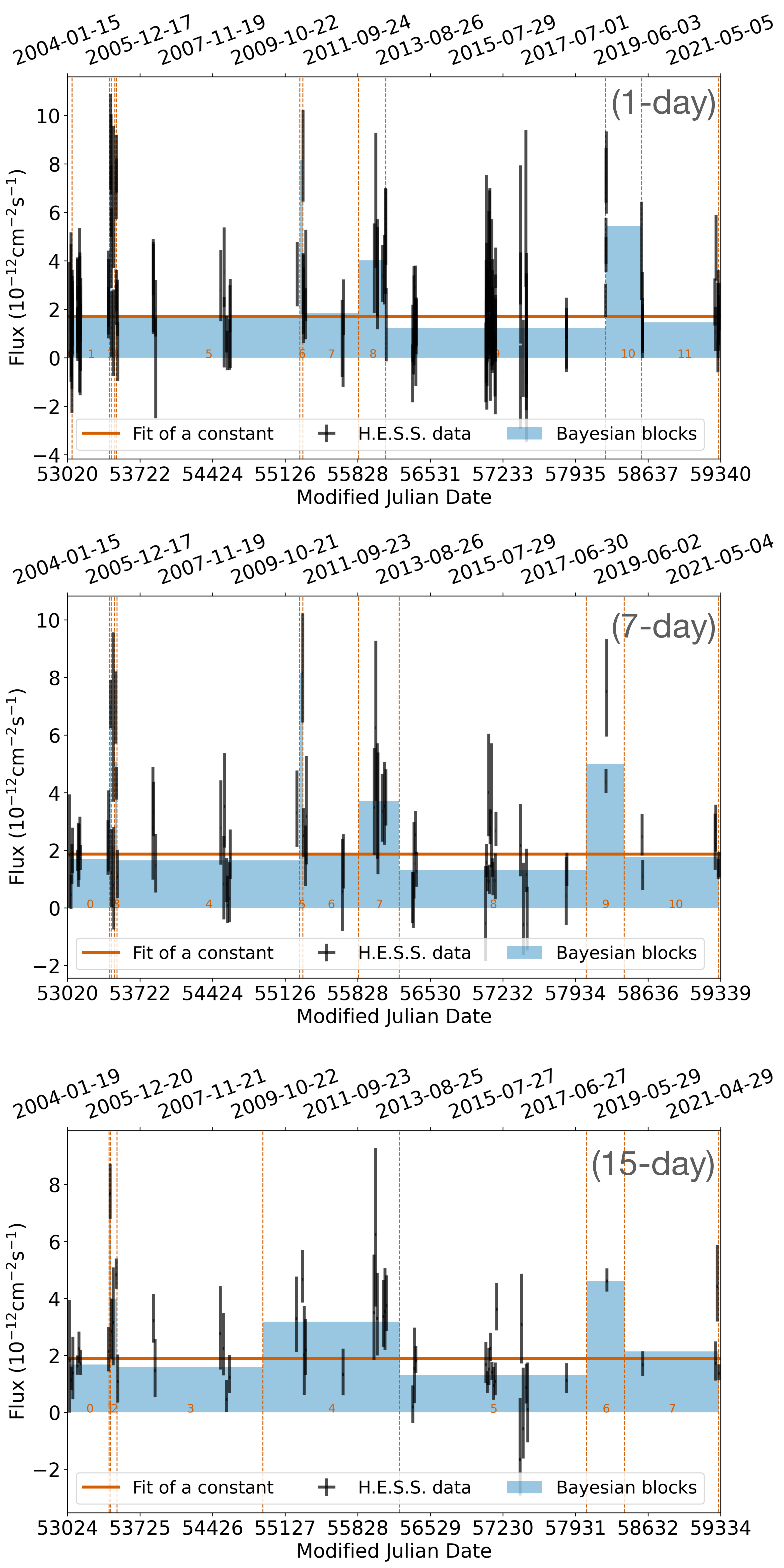}
    \caption{M\,87 long-term light curves with their Bayesian blocks derived according to Sect.~\ref{sec:bayesian_methods} for (top) one-day, (middle) 7-day, and (bottom) 15-day binned flux points. The H.E.S.S. data points are displayed by the black data points and a constant fit by the orange solid line. The blocks with their flux levels are given by the blue rectangles. Dashed orange lines indicate a change in the flux level. The orange numbers at the bottom indicate the labels of the blocks.}
    \label{fig:lightcurves}
\end{figure}

M\,87 is known to be highly variable in VHE gamma-rays, with flares of the order of one day \citep{2010flare}. Therefore, a daily-binned light curve could well identify the flares and isolate the low-state periods. Nevertheless, M\,87 is a rather weak source in the TeV regime at low states. In fact, at least 5 hours of observations are needed for a significant source detection \citep[see ][]{EHT2017}, hardly achievable in a single night. The choice of a daily-binned light curve would lead to large statistical uncertainties in the data points, and hence, an uncertain definition of the source states. On the other hand, the choice of very large bin sizes, for instance 60 days, would mix high and low states and result in a flattened light curve. Given the trade-off between reasonable statistical uncertainties in the flux points and a low state defined with the least variable emission in it, we chose the compromise of having a bin size equal to 30 days. The published flares from 2005 \citep{2005flare}, 2008 \citep{2008flare}, and 2010 \citep{2010flare} are well visible in the 30-day-binned light curve as shown in Fig.~\ref{fig:bayesian}. Furthermore, we also produced the 1, 7, and 15-day-binned light curves and their Bayesian blocks as shown in Fig.~\ref{fig:lightcurves}. In fact, the blocks with elevated flux in these light curves mostly coincided with the elevated blocks from the 30-day-binned light curve (Fig~\ref{fig:bayesian}). Nevertheless, very long blocks, for instance blocks 5 and 9 in the one-day light curve and blocks 4 and 8 in the 7-day light curve, show that the statistical uncertainties of the flux points are still too high for new flux levels to emerge from the Bayesian block analysis. More importantly, the VHE gamma-ray flare from 2008 falls within blocks below the average in the light curves of Fig.~\ref{fig:lightcurves}. Therefore, the light curves with bin sizes up to 15-day long have shown to be insufficient for a reliable Bayesian block analysis with M\,87 data.

\section{The interplay between CR pressure and gamma-ray emission}
\label{app:formalism}

Following the formalism from \citet{corecluster2}, a single beta profile function is utilized to parametrize the electron distribution as seen by Chandra X-Ray Observatory \citep{Chandra} in the inner Virgo Cluster:

\begin{equation}
n_\mathrm{e}(r) = n_\mathrm{0} \left [ 1 + (r/r_\mathrm{c})^2 \right ] ^{-3\beta/2},
\label{eq:n_e}
\end{equation}

\noindent where $n_\mathrm{0}$=0.230$\,$cm$^{-3}$, $\beta$=0.29 and $r_\mathrm{c}$=0.6$\,$kpc. The equation is valid for r<44$\,$kpc. For a fully ionized ICM composed of a hydrogen mass fraction of X=0.7 and helium mass fraction of Y=0.28, the ICM density distribution is given by:

\begin{equation}
n_\mathrm{ICM}(r) = \dfrac{\mu_\mathrm{e}}{\mu} n_\mathrm{e}(r),
\label{eq:n_ICM}
\end{equation}

\noindent where $\mu_\mathrm{e}$=1.18 and $\mu$=0.62 are the mean molecular weight per electron and per particle in the ICM, respectively.

The CRp energy and spatial distributions can be described as a function of the distance from the cluster center $r$ and the CRp dimensionless momentum $p_\mathrm{p}$:

\begin{equation}
n_\mathrm{p}(r, p_\mathrm{p}) = C_\mathrm{p}(r)p_\mathrm{p}^{-\alpha_\mathrm{p}}H(p_\mathrm{p}-q_\mathrm{p}),
\label{eq:n_CRp}
\end{equation}

\noindent where $p_\mathrm{p}=P_\mathrm{p}/(m_\mathrm{p} c)$, $P_\mathrm{p}$ is the CRp momentum, $m_\mathrm{p}$ the proton mass and $c$ the speed of light, $C_\mathrm{p}(r)$ is the spatial distribution of CRp, $H$ is the Heaviside step function, $q_\mathrm{p}=0.5$ is the dimensionless lower momentum cut-off and $\alpha_\mathrm{p}$ is the spectral index of the CRp distribution.

The gamma-ray source density distribution $s_\mathrm{\gamma}(E_\mathrm{\gamma},r)$ produced locally by the neutral pion decay can be derived based on the $n_\mathrm{ICM}(r)$ and $n_\mathrm{p}(r)$ distributions. The integration of $s_\mathrm{\gamma}(E_\mathrm{\gamma},r)$ from $E_\mathrm{1}$=300$\,$ GeV to $E_\mathrm{2}$=$\infty$, considering the $p$-$p$ cross-section $\sigma_\mathrm{pp}$ is given by:

\begin{equation}
\label{eq:5}
\begin{split}
&\lambda_{\gamma}(r) = \int_{300 \mathrm{GeV}}^{\infty}dE_{\gamma}s_{\gamma}(E_{\gamma},r) = \\
& \frac{4C_\mathrm{p}(r)}{3\alpha_\mathrm{p}\delta_\gamma} \frac{m_{\pi^0}c \sigma_{\mathrm{pp}}n_\mathrm{ICM}(r)}{m_p}
\left( \frac{m_\mathrm{p}}{2m_{\pi^0}}\right)^{\alpha_\mathrm{p}}\left [ B_\mathcal{X} \left( \frac{\alpha_\mathrm{p} + 1}{2\delta_\gamma},\frac{\alpha_\mathrm{p}-1}{2\delta_{\gamma}}\right) \right]_{\mathcal{X}_1}^{\mathcal{X}_2},\\
\end{split}
\end{equation}

\noindent in units of cm$^{-3}$s$^{-1}$, where $\sigma_\mathrm{pp}=3.2\cdot10^{-26}(0.96+e^{4.4-2.4\alpha_\mathrm{p}})\,$cm$^{-2}$ 
\citep{Pfrommer2004}, $m_\mathrm{\pi^0}$ is the neutral pion mass, $B_\mathcal{X}$(a,b) is the incomplete beta function, $\mathcal{X_\mathrm{i}}$ is given by:

\begin{equation}
\label{eq:6}
\mathcal{X_\mathrm{i}} = \left [ 1 + \left ( \dfrac{m_{\pi^0}c^2}{2E_\mathrm{\gamma,i}}\right )^{2\delta_\gamma} \right ]^{-1},
\end{equation}

\noindent $E_\mathrm{\gamma}$ is the gamma-ray energy calculated at i=$E_\mathrm{1}$ and i=$E_\mathrm{2}$, $\delta_\gamma \approx 0.14\alpha_\mathrm{p}^{-1.6} +0.44$ is the shape factor and $B_\mathcal{X}[(a,b)]_{\mathcal{X}_1}^{\mathcal{X}_2} = {B_\mathcal{X}}_2(a,b) - {B_\mathcal{X}}_1(a,b)$.

The uncertainty in the particle interaction model, specifically in the parametrization of the cross section $\sigma_\mathrm{pp}$, results in an uncertainty of approximately 50\% in the energy-integrated gamma-ray source function for energies ranging from 300 GeV to tens of TeV. Additionally, the systematic uncertainty associated with the detected gamma-ray flux with H.E.S.S. is approximately 20\% \citep{crabhess2006}. As a result, estimates on the CRp pressure and total energy in CRs should be considered order-of-magnitude estimates. As for the gamma-ray flux estimates, the CRp pressure $P_\mathrm{CR}$(r) depends on the CRp spatial distribution $C_\mathrm{p}$(r) and on the proton spectral index $\alpha_\mathrm{p}$:

\begin{equation}
\label{eq:7}
P_\mathrm{CR}(r) = \dfrac{1}{6}m_p c^2 C_\mathrm{p}(r) \left [ \mathcal{B}_{\left ( \dfrac{1}{1+q_p^2}\right ) }\left ( \dfrac{\alpha_p -2}{2},\dfrac{3-\alpha_p}{2}\right )\right ].
\end{equation}

The CR pressure distribution can be also represented as an energy density distribution, considering an effective adiabatic index for the CRs (fully relativistic value) of $\gamma_\mathrm{CR}$=4/3 \citep{corecluster1}:

\begin{equation}
P_\mathrm{CR}(r)=(\gamma_\mathrm{CR} - 1 ) e_\mathrm{CR}(r).
\end{equation}

The integral of $e_\mathrm{CR}(r)$ in the volume around the source gives the total energy in CRp in units of erg:

\begin{equation}
U_\mathrm{CR} = \int_V e_\mathrm{CR}(r) dV.
\end{equation}

The representation of the CR pressure $P_\mathrm{CR}(r)$ in terms of the thermal pressure $P_\mathrm{th.}(r)$ is useful to characterize the steady state model, in which the heating of the ICM counterbalances the cooling:

\begin{equation}
\label{eq:10}
X_\mathrm{CR}(r) = \dfrac{P_\mathrm{CR}(r)}{P_\mathrm{th.}(r)},
\end{equation}

The thermal pressure $P_\mathrm{th.}(r)$ is given by

\begin{equation}
\label{eq:11}
P_\mathrm{th.}(r)=\dfrac{\mu_e}{\mu} n_\mathrm{e}(r)k_\mathrm{B}T(r),
\end{equation}

\noindent where $k_\mathrm{B}$ is the Boltzmann constant and T(r) is the temperature profile of the X-ray emitting electrons. As the density distribution of the electrons in the plasma $n_\mathrm{e}(r)$, T(r) is also parametrized from X-ray Chandra data \citep{corecluster1}:

\begin{equation}
\label{eq:12}
T(r)= T_\mathrm{0} + (T_\mathrm{1} - T_\mathrm{0}) \left [1 + \left ( \dfrac{r}{r_\mathrm{T}} \right )^{-\eta} \right ]^{-1} \left [ 1 + \left ( \dfrac{r}{ar_\mathrm{200}} \right )^2 \right ]^{-0.32},
\end{equation}

\noindent where $T_\mathrm{0}$=1.9$\,$keV, $T_\mathrm{1}$=3.1$\,$keV, $r_\mathrm{t}$=28$\,$kpc, $\eta$=1.4, a=0.2 and $r_\mathrm{200}$=1.08$\,$Mpc.

\section{The diffuse emission templates}
\subsection{The magnetic confinement template}
\label{app:lofar}

\begin{figure}[ht!]
\centering
\includegraphics[width=\textwidth/2]{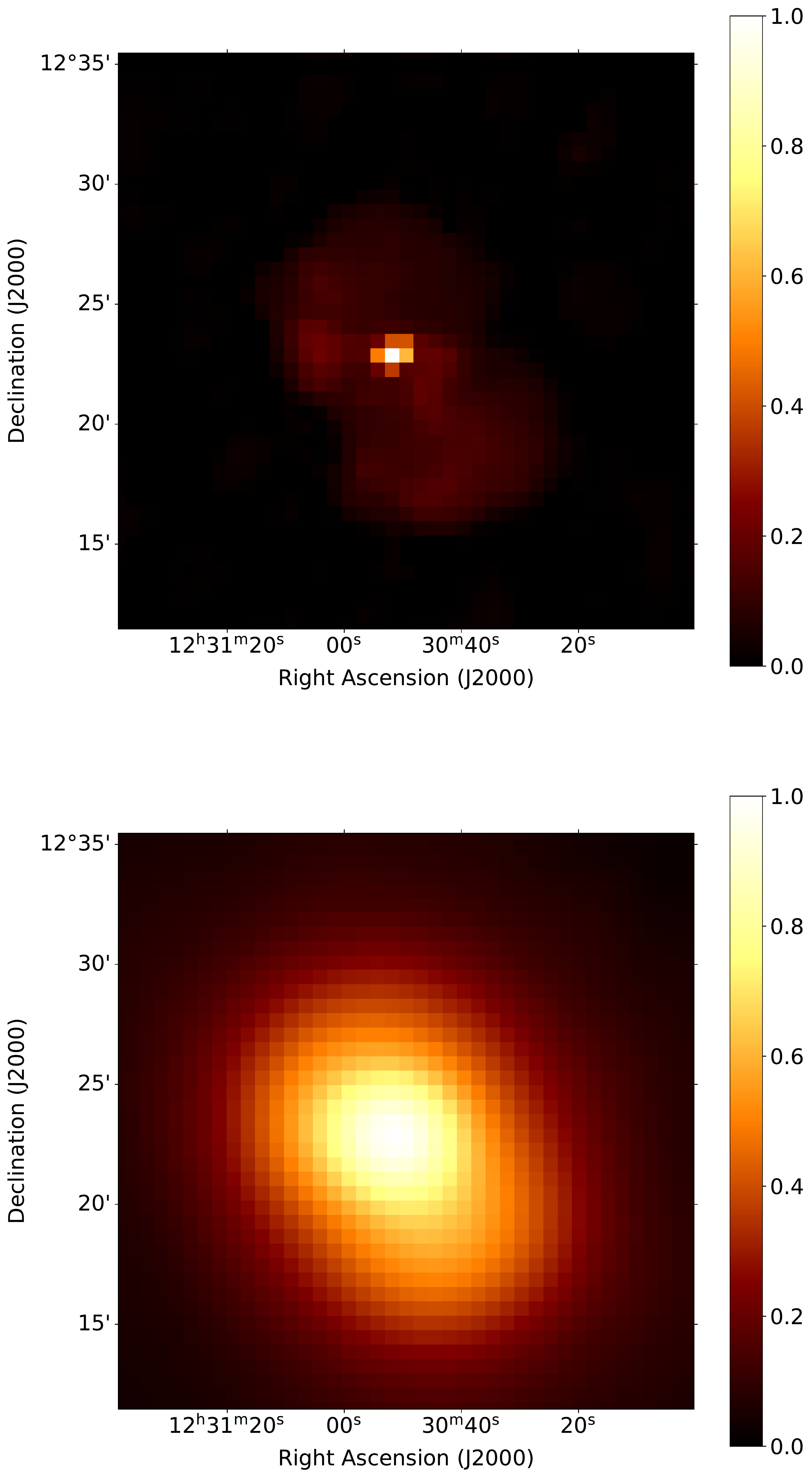}
\caption[H.E.S.S. resolution of the 2D templates for the gamma rays emitted in the source due to $\pi^0$ decay.]{2D templates for the gamma-ray emission in the inner Virgo Cluster, following the square root of the radio emission intensity normalized at its maximum as seen by LOFAR (Fig. 5 from \citet{LOFAR}) for better visualization: (top) the model displayed with a bin size of $\ang{;;36}$ and (bottom) the intrinsic emission convolved with the H.E.S.S. PSF.}
\label{fig:LOFAR}
\end{figure}

We consider the LOw Frequency Array \citep[LOFAR, ][]{LOFAR_intro} study of M\,87. The 140$\,$MHz radio emission \citep{LOFAR} traces relativistic electrons, which emit synchrotron photons giving rise to the $\approx$30$\,$kpc micro-halos \citep{corecluster1}. The radio micro-halos are very well confined within boundaries that have the same dimensions at all radio frequencies down to 25$\,$MHz \citep{LOFAR}. This indicates that the distribution of non-thermal electrons is energy independent and that they are magnetically confined in the lobes. CRp accelerated in the central AGN populate the cluster and could also be present up to the same boundaries as seen in the radio band. Apart from X-ray-depleted regions (cavities) in radio galaxies, where the composition of the material is yet unknown \citep{Abdulla2019}, the ICM at the inner Virgo cluster has a rather high density of 0.1-0.01$\,$cm$^{-3}$ \citep{corecluster1}. Therefore, assuming that the CRp mix with the ICM, the CRp could hadronically interact with it and produce pions. While the charged pions decay to electrons and positrons which likely contribute to part of the radio synchrotron emission from the micro halo, the neutral pions decay to gamma rays. The morphology of the hadronic gamma-ray emission depends not only on the ICM density distribution but also on the CRp energy and spatial distributions (Appendix \ref{app:formalism}). A complete model of this emission would demand a large number of assumptions, for instance, on the poorly known content of the X-ray cavities and the distribution of CRp in the cluster. Therefore, for the sake of simplicity, we assume in this approach that the gamma rays follow the same spatial distribution as the radio emission detected by LOFAR at 140$\,$MHz. This allows us to probe the contribution of an extended hadronic component to M\,87 low-state emission despite the different radiation mechanisms. Based on this assumption, we generated a 2D template for the gamma-ray emission, the magnetic confinement template, shown in Fig.~\ref{fig:LOFAR} as it would be seen with H.E.S.S.

Naturally, we do not expect that the magnetic confinement template explains the entirety of M\,87 low-state VHE emission because the template does not account for the emission from the AGN and this study has not measured an extension in the Gaussian model (Sect.~\ref{sec:morphology_results}).

Although we utilized the template shown in Fig.\ref{fig:LOFAR} for the main analysis in Sect.\ref{sec:discussion}, we also tested an alternative template. This alternative template attempted to reduce the contribution of the kpc-jet to the overall shape of the emission by masking the central bin in Fig.~\ref{fig:LOFAR} (top) before convolving it with the H.E.S.S. PSF. We found that the morphology fit of the alternative template showed a maximum allowed contribution of 30\% to the total low state gamma-ray emission of M87 ($\lesssim$4.5$\times$10$^{-13}\,\mathrm{cm^{-2}s^{-1}}$), compared to 45\% when using the original template. However, we opted to proceed with the original approach since it yields a more conservative UL on the CR pressure.

\subsection{The steady-state template}
\label{app:steady}

\begin{figure}[ht!]
\centering
\includegraphics[width=\textwidth/2]{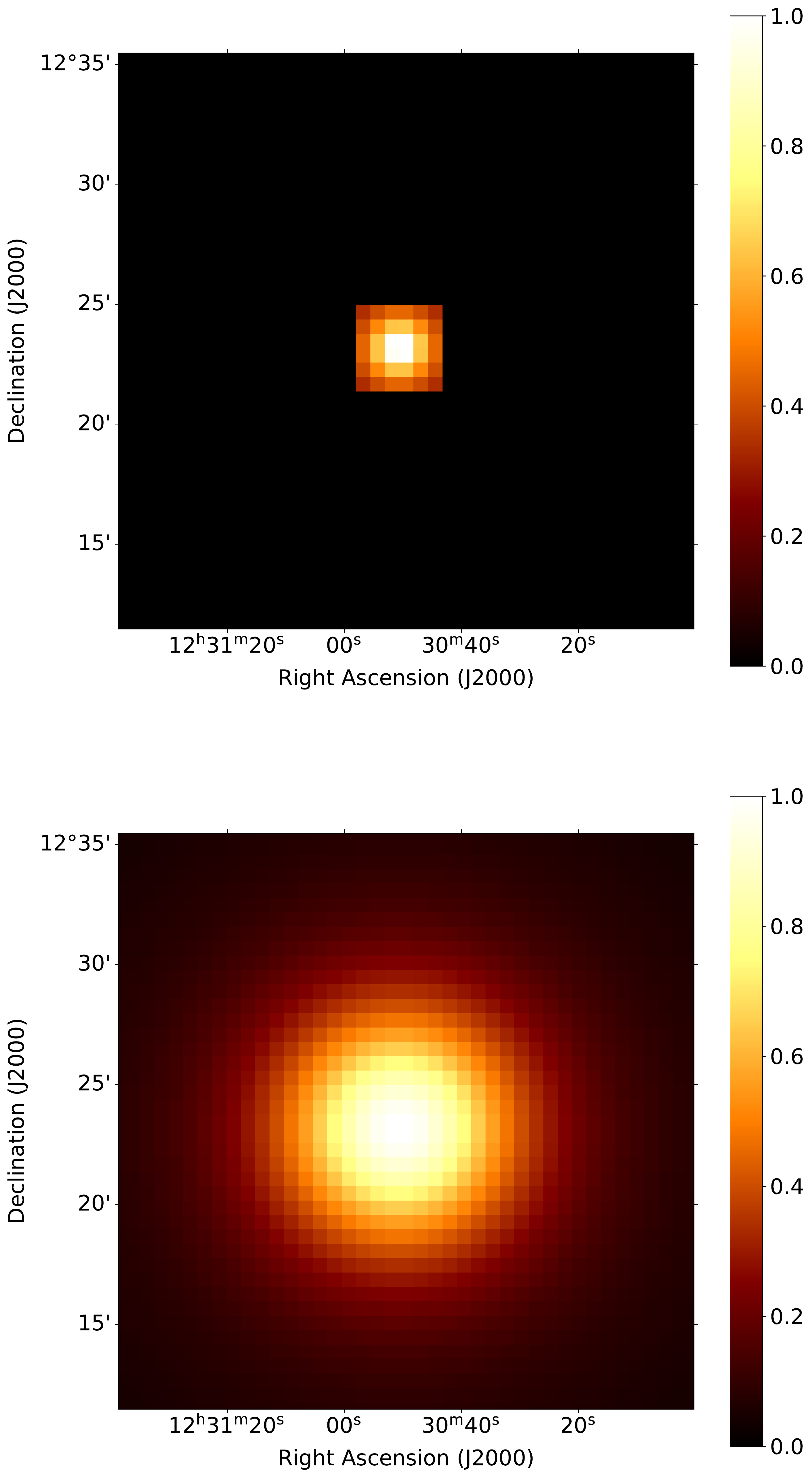}
\caption{2D templates for the gamma-ray emission in the inner Virgo Cluster, following the square root of the gamma-ray intensity due to $\pi^0$ decay considering the angular resolution of the H.E.S.S. sky maps: (top) the model displayed with a bin size of $\ang{;;36}$ and (bottom) the intrinsic emission convolved with the H.E.S.S. PSF.}
\label{fig:radio_cluster}
\end{figure}

In the second approach, we adopted the model from \citet{corecluster1,corecluster2} for the AGN feedback. The authors propose a steady-state model, in which the CRp pressure is sufficient to counterbalance, at every distance from the cluster center, the radiative cooling of the ICM. The CRp excite Alfvén waves through the streaming instability and the non-linear Landau damping of these waves provides an efficient mechanism for heating the ICM. In addition to the CR pressure, the thermal pressure helps halt the cooling flow toward the cluster center. The model solves the cooling flow problem and predicts a steady and extended gamma-ray signal due to neutral pion decay. The assumption of a steady state leads to an analytical estimation of the CRp pressure distribution in the inner cluster. Using the CRp pressure distribution in Fig.~A.1 from \citet{corecluster1} and equations from Appendix~\ref{app:formalism} we derived the distribution of gamma rays above 300$\,$GeV produced by the decay of neutral pions for the same set of CRp distributions utilized in the first approach of our study. Afterwards, we integrated the gamma-ray distribution along the line of sight to generate a second 2D gamma-ray emission template as it would be seen with H.E.S.S. The template is shown in Fig.~\ref{fig:radio_cluster}. We also integrated the gamma-ray distribution in the volume around the source ($\lesssim$20$\,$kpc) and accounted for the distance to M\,87 to obtain a predicted VHE gamma-ray flux of $2.2\times10^{-13}$cm$^{-2}$s$^{-1}$ by the steady-state model with $\alpha_\mathrm{p}$=2.1, that is $\approx$15\% of the low state of M\,87 VHE emission (Table \ref{table:statesmaster}). The predicted flux becomes smaller for steeper indices and is, therefore, less relevant for this study.

\end{appendix}

\end{document}